\documentclass[12pt,referee,oneside,reqno]{article}

\usepackage{graphicx}
\usepackage{amssymb}
\usepackage{epstopdf}
\usepackage{setspace}
\usepackage{natbib}
\usepackage{algorithmic}
\usepackage{xr}
\usepackage[top=1.45in, bottom=1.3in, left=.876in, right=.876in]{geometry}

\usepackage{placeins}

\usepackage{rotating}
\usepackage{amsmath,amsthm}
 \numberwithin{equation}{section}
 
\newtheorem{theorem}{Theorem}[section]
\newtheorem{lemma}[theorem]{Lemma}

\usepackage{mathtools}
\usepackage{algorithmic}
\usepackage[ruled]{algorithm}
\usepackage{color, colortbl}

\definecolor{Tgray}{rgb}{.9,.9,.9}

\DeclareMathOperator{\y}{\mathbf{y}}
\DeclareMathOperator{\x}{\mathbf{x}}
\DeclareMathOperator{\cc}{\mathbf{c}}
\DeclareMathOperator{\vv}{\mathbf{v}}

\DeclareMathOperator{\rr}{\mathbf{r}}
\DeclareMathOperator{\z}{\mathbf{z}}
\DeclareMathOperator{\uu}{\mathbf{u}}

\DeclareMathOperator{\q}{\mathbf{q}}

\DeclareMathOperator{\s}{\mathbf{s}}

\DeclareMathOperator{\0}{\mathbf{0}}

\DeclareMathOperator{\X}{\mathbf{X}}
\DeclareMathOperator{\B}{\mathbf{B}}
\DeclareMathOperator{\C}{\mathbf{C}}
\DeclareMathOperator{\D}{\mathbf{D}}
\DeclareMathOperator{\G}{\mathbf{G}}

\DeclareMathOperator{\U}{\mathbf{U}}
\DeclareMathOperator{\V}{\mathbf{V}}
\DeclareMathOperator{\Y}{\mathbf{Y}}

\DeclareMathOperator{\Pb}{\mathbf{P}}

\DeclareMathOperator{\Z}{\mathbf{Z}}
\DeclareMathOperator{\HH}{\mathbf{H}}
\DeclareMathOperator{\Sb}{\mathbf{S}}

\DeclareMathOperator{\I}{\mathbf{I}}
\DeclareMathOperator{\A}{\mathbf{A}}
\DeclareMathOperator{\W}{\mathbf{W}}

\DeclareMathOperator{\bbeta}{\boldsymbol{\beta}}
\DeclareMathOperator{\bmu}{\boldsymbol{\mu}}

\DeclareMathOperator{\blambda}{\boldsymbol{\lambda}}

\DeclareMathOperator{\balpha}{\boldsymbol{\alpha}}
\DeclareMathOperator{\Omeg}{\boldsymbol{\Omega}}

\DeclareMathOperator{\rrho}{\boldsymbol{\rho}}

\DeclareMathOperator{\Xcal}{\boldsymbol{\mathcal{X}}}

\DeclareMathOperator*{\argmin}{arg\,min}

\title{\vspace*{-.1in}\large  \bf Local-Aggregate Modeling for  Big-Data via Distributed Optimization: Applications to Neuroimaging}
\author{{ \bf \small Yue Hu$^{*}$} \\ {\footnotesize Department of Statistics, Rice University}\\
{\footnotesize $^{*}$email: yue.hu@rice.edu} \\{\bf \small and }\\ {\small \bf Genevera I. Allen$^{*}$}\\
{\footnotesize Dobelman Family Junior Chair}\\
{\footnotesize Departments of Statistics and Electrical \& Computer Engineering, Rice
University, }\\
{\footnotesize Department of Pediatrics-Neurology, Baylor College of Medicine,} \\
{\footnotesize Jan and Dan Duncan Neurological Research Institute, Texas Children's
Hospital.}\\ {\footnotesize $^{*}$email: gallen@rice.edu}
}
\date{}
\begin{document}
\maketitle
\vspace*{-.1in}
\noindent{\bf \footnotesize Summary: }{\footnotesize Technological advances have led to a proliferation of 
structured big data that have matrix-valued covariates. We are specifically motivated
to build predictive models for multi-subject neuroimaging data based
on each subject's brain imaging scans.  This is an
ultra-high-dimensional problem that consists of a matrix of covariates
(brain locations by time points) for each subject; few methods
currently exist to fit supervised models directly to this tensor data.
We propose a
novel modeling and algorithmic strategy to apply generalized linear
models (GLMs) to this massive tensor data in which one set of
variables is associated with locations.  Our method
begins by fitting GLMs to each location separately, and then builds an ensemble by
blending information across locations through regularization with what we term
an aggregating penalty. Our so called, Local-Aggregate Model, can be
fit in a completely distributed manner over the locations using an Alternating Direction Method of Multipliers (ADMM) strategy, and thus
greatly reduces the computational burden. Furthermore, we propose to
select the appropriate model through a novel sequence of faster algorithmic
solutions that is similar to regularization paths.  We will demonstrate
both the 
computational and predictive modeling advantages of our methods via
simulations and an EEG classification problem. }

\newpage

\section{\bf Introduction}
\label{sec:intro}

Predictive modeling of subject-level behavior based on whole-brain
scans is an important goal of neuroimaging studies.  Technologies such
as EEG, MEG, and fMRI produce massive amounts of
spatio-temporal data (brain images or recordings measured over time)
for each subject that serve as covaraites; but most
studies typically consist of only tens to hundreds of subjects from
which to build supervised predictive models. 
This problem is thus ultra-high-dimensional and poses
both methodological and computational 
problems for classical statistical and machine learning techniques.
First, in $p\gg n$ settings, 
classical statistical models are ill-posed and cannot be fit without 
some form of regularization, dimension reduction, or variable
selection. 
Second, the predictors in this supervised learning problem are not vectors as is typical, but are
matrices; in other words, our data can be arranged as a 3D array of 
subjects by brain locations by time points.  Third, the size of this
big data (for example, $\approx$4GB per subject for fMRI data) can
easily exceed the memory limit of mathematical 
programming tools or even the capacity of a single computer node.  
Finally, the covariates for each subject are spatio-temporal, and few 
existing statistical learning methods
directly account for these strong structural dependencies.   In this
paper, our objective is to develop a novel multivariate modeling
framework 
and fast algorithmic strategy for multi-subject neuroimaging studies
that 
addresses each of the above challenges and leads to 
better predictive performance with scientifically more interpretable
results.

Many statistical machine learning techniques have been proposed
for supervised modeling of multi-subject neuroimaging data.  
First, one could apply standard
multivariate machine learning techniques to whole-brain data
\citep{pereira2009machine}, such as support vector machines \citep{de2007classification},
which typically take a vector of
covariates as inputs.  
To apply these machine learning methods then, the locations
by time series for each subject must be vectorized, thus unraveling
the important 
spatio-temporal data structure. Alternatively,
one could apply machine learning 
techniques \citep{de2007classification,calhoun2009review} after first reducing the dimension of each subject's brain
scan using methods such as principal components or independent
components analysis.
While dimension reduction
techniques effectively reduce the computational burden of
multi-subject modeling, the resulting predictions are not directly
interpretable in the original domain.  
Recently, there has been some interest in fitting supervised models to
a tensor of covariates in the statistical literature.  The
methods of \citep{hung2013matrix,zhou2013tensor,zhou2013regularized}
all enforce some form of rank constraints on matrix or tensor coefficients.
While low-rank models are a methodological solution to fitting
multi-subject models in ultra-high-dimensional settings, these methods
are computationally intensive; for example, the method of
~\citep{zhou2013regularized} requires computing repeated singular
value decompositions.  Furthermore, none of these aforementioned
approaches or standard machine learning methods directly take the
strong spatio-temporal dependencies 
observed in neuroimaging data into account.  We seek a method that
both naturally deals with a tensor of covariates and also respects the
spatio-temporal nature of whole-brain images.

The development of our modeling strategy will be primarily motivated by
computational concerns.
Fitting statistical learning models to ultra-high-dimensional data
such as whole-brain multi-subject neuroimaging data is a major
computational hurdle requiring huge amounts of memory ($\approx$ 4GB
per subject) and long
processing times.  A possible way around this is to break up the
multivariate problem into a series of smaller problems that can be fit
independently.  One could imagine fitting a separate statistical model
to each brain location (e.g. voxel); for each of these ``local''
models, we are back in the common framework of a vector of
covariates, the time series, for each subject.  
This has close 
connections with the ``massive univariate analysis'' such as the
random effects general linear model~\citep{friston1994statistical} that is widely used for finding spatial maps of brain activation in fMRI data.  
Such massive
independent analyses can be computed efficiently by using parallel
computing such as with Message Passing
Interfaces (MPIs) or Graphic Processing Units (GPUs).  While these massive
univariate frameworks are appealing because their computation can be
distributed, they do not result in a unified whole-brain model to predict
subject-level attributes.
Furthermore, we expect adjacent brain locations
to be highly correlated.  Massively univariate methods then
lose important information that can be gained by considering all
brain locations through a multivariate model.

We seek to develop a method that has the computational advantages of
massively univariate methods, but still directly accounts for the
spatio-temporal tensor structure of multi-subject neuroimaging
studies.  To this end, we develop a framework that fits
separate models to each brain location, but then combines these set of local models in an ensemble that blends information across nearby local models through regularization.

In this paper, we make several methodological and computational
contributions.  Methodologically, we introduce a novel modeling
framework for (1) tensor-valued data as (2) an ensemble of local predictions that (3) directly incorporates
spatial and temporal information through regularization.  As
previously mentioned, this framework fits separate models to each
location like the massive univariate frameworks, but then blends
information across locations using regularization.
Our overall ensemble is then the aggregate or sum of all the individual local
models with the regularization term serving to both smooth local model
coefficients spatially and weight the local models according to their
predictive ability.  While our so-called {\em Local-Aggregate}
modeling framework (Section~\ref{sec:model}) is general, in this paper, we
specifically focus on 
predictive modeling via Generalized Linear Models (GLMs).  Because of
the unique structure of our Local-Aggregate model, we can (4) fit this
multivariate model in a fully distributed manner (both in terms of computation and more importantly, memory and data storage) via a simple
splitting algorithm, the alternating direction method of multipliers (ADMM), 
 a major computational contribution (Section~\ref{sec:algorithm}).
Additionally, our algorithm leads to (5) a novel strategy to speed the
model selection process by using what we term an
{\em algorithm path} as an approximation to computationally involved
regularization paths (Section~\ref{sec:alg_path}).   We will show through
simulations (Section~\ref{sec:sim}) and a real multi-subject EEG
classification problem (Section~\ref{sec:eeg}) that 
our methodological contributions lead to a model with improved
predictive performance as well as scientifically more interpretable
results while our computational contributions can dramatically
reduce the computational burden of fitting ultra-high-dimensional
models.

\section{\bf Methods: Local-Aggregate Modeling}
\label{sec:model}

Before we begin, let us review the notation we will use in this paper.
Tensors are denoted by $\Xcal$, matrices by $\X$, vectors by $\x$ and scalars by
$x$.  It maybe necessary to vectorize a matrix $\X \in \Re^{n \times p}$ as $\mathrm{vec}(\X)\in\Re^{np}$; or
matricize a tensor along a particular mode: for $\Xcal$ of size $n
\times p \times q$, $\X_{(1)}$ is of size $n \times pq$.  Outer
products of two vectors will be denoted by $\x \circ \y$,
and Kronecker products by $\X \otimes \Y$. 
The $\ell_2$ norm over groups, $g\in \mathcal{G}$, is given by $\sum_{g\in
  \mathcal{G}}\|\x_g\|_2$.
  
Our goal is to develop a modeling framework for predicting
subject-level responses based on a three-dimensional tensor array of
covariates corresponding to a matrix of predictors for each subject.
As we are primarily motivated by modeling multi-subject neuroimaging
data such as with EEG and fMRI, we denote our subject-level response
as $\y \in \Re^{n}$ for continuous behavioral outcomes or $ \y \in \{0,1\}^n$ for binary disease categories, and 
our predictors as $\Xcal_{n\times \tau \times L}$ for $n$ independent
subjects, $\tau$ time points and $L$ brain locations.  
Our primary considerations in developing a new multi-subject predictive model are
to directly account for the tensor structure of the predictors, the
ultra-high-dimensionality of the problem, and the spatio-temporal
nature of neuroimaging data.  We also seek to structure our model so
that estimation is computationally scalable for big data.

\subsection{Local Models}
\label{sec:local}
We assume that the generative model for the pair of responses and
tensor covariates, $(\y, \Xcal)$, follows a matrix generalized linear
model (GLM): $g(\bmu) = \alpha + \X_{(1)}^T vec(\B)$, where $\bmu
= \mathbb{E}(\y|\Xcal)$ is the conditional mean responses, $g()$
is the 
canonical link function associated with a particular exponential 
family (e.g. the identity link, $g(\mu) = \mu$ for the Gaussian
distribution), $\alpha \in \Re$ is the intercept, and the coefficient
matrix $\B \in \Re^{\tau \times L} = [\bbeta_1 \ldots \bbeta_L]$ is
the collection of local coefficient vectors $\bbeta_l \in \Re^\tau$. 
This multivariate model follows the construction of classical
GLMs \citep{mccullagh1984generalized} treating the vectorized
locations by time points as covariates.

These matrix GLMs are ultra-high-dimensional statistical problems with
$n$ in the tens to hundreds of subjects relative to $p = L\tau$ on the
order of 
tens of thousands to millions.  Hence, several statisticians have
recently proposed regularization methods to fit matrix GLMs by using
low-rank constraints, penalties, or low-rank structured coefficient
models \citep{zhou2013regularized,zhou2013tensor} or by
placing two-way penalties on the coefficient 
matrix \citep{tian2012two}.  These matrix GLM methods,
however, suffer from both computational and statistical
inefficiencies.  Even with regularization, the number of parameters to
be estimated is huge relative to the number of independent
observations and thus we cannot expect to accurately estimate matrix
GLM parameters.  Second, matrix GLMs cannot be fit in an easily
distributable manner. Consider a simple linear model, for example, with
squared error loss: $\| \y - \X_{(1)}\mathrm{vec}(\B) \|_2^2 = \sum_{i
= 1}^n (y_i - \sum_{t = 1}^\tau \sum_{l = 1}^L x_{itl} \beta_{tl}
)^2$; as the matrix parameters are coupled together in
the loss function, it is not conducive to distributed optimization.

Thus, we are motivated to consider an alternative way to model the
responses and tensor covariates that will approximate the matrix GLM
but also be statistically and computationally more efficient.  We
consider modeling the response at each location separately through
local GLMs: $ g(\boldsymbol{\mu}_l)  = \alpha_l + \X_l
\bbeta_l , \forall \ l=1, \ldots L$. 
Here, $\X_l \in \Re^{n\times \tau}$ is data at location $l$,
$\boldsymbol{\mu}_l = \mathbb{E} (\y | \X_l)$ is the local conditional
mean, and $\alpha_l \in \Re$ is the local intercept.   We propose to
build an ensemble of these local models by aggregating through 
regularization, which we will discuss in the next section.  Notice
first, however, that the coefficients in our collection of local
GLMs are clearly only equivalent to the coefficients of the matrix GLM
if each location is independent.  But with neuroimaging data, we
expect brain locations to be strongly spatially correlated.  A key
modeling assumption, 
then, is that if we account for the spatial dependencies of the
parameters (through regularization, for example, discussed
subsequently), then the local models are approximately independent.

If our proposed collection of local models are only an
approximation to the true
matrix GLM, then why do we expect these models to work well in
practice?  We answer this question fully in Section~\ref{sec:advantage} after we
introduce our optimization framework.  Briefly, however, notice that
each of our local models has $\tau$ parameters, far fewer relative to
the matrix GLM.  Thus, we expect our statistical efficiency in
estimating parameters of the local model to be much better than that
of the matrix GLM.  Second, fitting our models is conducive to simple
distributed optimization.  Consider again the example of squared error
loss, now for the sum of our local models: $\sum_{l=1}^L \| \y
- \X_l \bbeta_l\|_2^2 = \sum_{i=1}^n \sum_{l=1}^L (y_i
- \sum_{t=1}^\tau x_{itl}\beta_{tl})^2$; this loss is
location-separable and hence easily distributable across locations.
Finally, our collection of local models can be viewed as an ensemble learning method, and ensembles are known to yield powerful predictive models.

\subsection{Local-Aggregate Optimization Framework}
\label{sec:general}

We propose to build an ensemble learning method by aggregating our
local models through regularization.  
Fitting a simple sum of local models, however, does not account for
the spatial structure of 
neuroimaging data.  Thus, we seek a method of combining our local
models in such a way to borrow strength across neighboring locations
through an {\it aggregating penalty}, $P_{agg}(\B,\G)$ with external
spatial information matrix $\G$ discussed subsequently. 
Also in many applications including neuroimaging, each of the local
GLMs could be high-dimensional if $\tau > n$, and thus, some local
regularization may also be needed.   
Then, our general Local-Aggregate optimization framework
below is an ensemble of regularized local GLMs with an
aggregating penalty: 
{\begin{equation}
\label{loc_agg_gen}
 \underset{\balpha,\B}{\text{minimize}}
\quad  \underbrace{ \sum_{l = 1}^L [ \underbrace{\ell(\y; \alpha_{l} +
      \X_{l} \bbeta_{l})}_{\text{GLM loss}}+
    \underbrace{\lambda_{loc} P_{loc}(\bbeta_{l})}_{\text{Possible local penalties}} ]}_{\text{Sum of penalized log-likelihoods for local GLMs}} 
\underbrace{+ \lambda_{agg} P_{agg}(\B,\G).}_{\text{Aggregating 
    over locations}} 
\end{equation}
}
\noindent Here, $\ell()$ is the GLM loss function, or the negative
log-likelihood, $P_{loc}(\bbeta_l): \Re^{\tau} \rightarrow \Re^{+} $
is a local penalty, $P_{agg}(\B,\G): (\Re^{\tau \times L}, \Re^{\tau\times\tau}) \rightarrow \Re^{+}$ is the aggregating penalty, and   
$\lambda_{loc} \geq 0, \lambda_{agg}\geq 0$ are tuning parameters of local and aggregating regularization
respectively.

\subsubsection{Aggregating Penalty}
\label{sec:agg_pen}

To account for the spatial dependencies in the data, we propose to
enforce the local coefficients to be smooth over locations through
our aggregating penalty.   
Popular approaches to achieve spatial smoothness include 
imposing smoothness with respect to a graph structure through
the graph penalty, $||\bbeta||_{\G}^2
= \bbeta^T \G \bbeta$ \citep{grosenick2013interpretable,allen2014generalized}.
Taking $\G$ to be the graph Laplacian (the difference between
the degree matrix $\D\in \Re^{L\times L}$ and the adjacency matrix $\W \in \Re^{L\times L}$, where $\W$ is a symmetric matrix with nonnegative entries $w_{l,l'}$ and $\D_{l,l'} = \sum_{l=1}^L w_{l,l'}$),  the graph
penalty has the appealing interpretation of regularizing pairwise
differences between coefficients that are adjacent in the graph:
$|| \bbeta||_{\G}^2 = \sum_{(i,j) \in \G}( \beta_i
- \beta_j)^2$ \citep{grosenick2013interpretable}.   
Consider a multivariate extension of a weighted 
graph Laplacian penalty where we penalize the weighted differences
between coefficient vectors: $\sum_{l \neq l'} w_{l,
l'}|| \bbeta_{\cdot l} - \bbeta_{\cdot l'}||_2^2$, where
$\bbeta_{\cdot l} \in \Re^{\tau}$ is the $l^{th}$ column of $\B$, and $w_{l,l'}$'s are weights inversely proportional to the distance between
locations $l$ and $l'$ (in
subsequent sections, we will use $\bbeta_{l}$ to
denote $\bbeta_{\cdot l}$ for notational convenience).
Note that if we take $\G$ to the be the $L \times L$ weighted graph
Laplacian, $\G = \D - \W$,
then the
multivariate graph 
Laplacian penalty is equal to $tr( \B \G \B^{T})$, i.e.,
$ tr( \B \G \B^{T}) =  \sum_{(l \neq l')} w_{l,
l'}|| \bbeta_{\cdot l} - \bbeta_{\cdot l'}||_2^2 $.

Hence, we can use an aggregating penalty of the form $P_{agg}(\B,\G)
= \mathrm{tr}(\B\G\B^T)$, 
noting that we can specify $\G$, and more
specifically the weights $w_{l,l'}$ defining $\G$, in such a way as to
interpret our penalty as a graph type penalty. Moreover,
as $ \sum_{(l \neq l')} w_{l,
l'}|| \bbeta_{\cdot l} - \bbeta_{\cdot l'}||_2^2  = \sum_{t=1}^\tau \bbeta_{t\cdot} ^T\G \bbeta_{t\cdot} $, we can also interpret $\G$ as a roughness penalty matrix often used for smoothing in functional data analysis \citep{ramsay2006functional}.

We are then left with the task of choosing the weights $w_{l,l'}$,
which is application
specific, but there are two general classes of well studied and
employed weighting schemes:  

$\bullet \quad w_{l,l'} = \left\{ 
  \begin{array}{l l}
    1 & \quad \text{if } l' \in \mathcal{N}(l)\\
   0 & \quad \text{otherwise}
  \end{array}\right.$, where $\mathcal{N}(l)$ is the neighborhood of node $l$.  This has been employed when locations are laid out on a regular grid as in fMRI data \citep{grosenick2013interpretable}, or when locations correspond to an existing network structure  \citep{huang2009analysis}.
  
$\bullet \quad w_{l,l'} = Kern(D_{l,l'}, \theta)$, where $Kern()$ is a  kernel smoothing function that takes $D_{l,l'}$, the distance between locations $l$ and $l'$.  While there are many examples of appropriate kernel smoothers \citep{ramsay2006functional}, a common smoother is the exponential kernel, $w_{l,l'} = e^{-D_{l,l'}^2/\theta}$.  These types of weights are useful for data where the locations correspond to some irregular physical coordinates, such as channels of EEG or MEG signals.

\subsubsection{Example: Local-Aggregate Modeling for Multi-Subject Neuroimaging Data}
\label{sec:loc_egg}
We pause to discuss some possible forms our Local-Aggregate modeling
 framework may take for our primary motivating example of
 multi-subject spatio-temporal neuroimaging data such as with
 EEG, MEG, or fMRI data. 
 As measurements are taken over time in these examples, we also expect
 the coefficients to be smooth with respect to time.  Therefore, we
 may wish to encourage temporal smoothness through  a local quadratic
 smoothing penalty $P_{loc}^{sm}(\bbeta_l)
 = \bbeta_l^T\Omeg \bbeta_l$. 
 where $\Omeg$ is the second order difference matrix penalizing the squared difference between coefficients at adjacent time points \citep{eilers1996flexible}.  
 Additionally, some neuroimaging data consists of  many locations; consider EEG data which can have up to 512 channels, not all of which are expected to contribute to subject-level responses.  Hence, we may also want to encourage sparsity to select only the relevant locations for prediction purposes.  However, since our local coefficient, $\bbeta_l$, is a vector, we cannot simply apply a scalar penalty, such as the $\ell_1$ norm.   Thus, we suggest a group lasso penalty, $\sum_{l=1}^L \| \bbeta_l\|_2$ \citep{yuan2006model}, which treats coefficients at each location as a group, either zeroing out all elements of $\bbeta_l$ or letting all elements be non-zero.  Also, this group lasso penalty can easily be localized,  $P_{loc}^{sp}(\bbeta_l)=\| \bbeta_l \|_2$, as it is separable in locations.  Putting everything together, we have the Local-Aggregate optimization framework for neuroimaging data:
{ \begin{equation}
\label{loss_neuro}
\underset{\balpha,\B}{\text{minimize}} \quad \sum_{l=1}^L \left[\ell(\y; \alpha_{l} + \X_{l} \bbeta_{l}) +
  \lambda_{loc}^{sm}\bbeta_l^T\Omeg \bbeta_l+  \lambda_{loc}^{sp}\|
  \bbeta_l \|_2 \right]  +   \lambda_{agg}\text{tr} (\B\G\B^T), 
\end{equation}}
where $\lambda_{loc}^{sm} \geq 0$ and $\lambda_{loc}^{sp} \geq 0$ are local tuning parameters, and the choice of $\G$ depends on the spatial information of the specific imaging modality as discussed in Section~\ref{sec:agg_pen}.

 \subsubsection{ Subject-level Behavior Prediction}

Once we obtain the estimates of all parameters, we propose to predict
the subject-level response by 
taking the average of the local predictions: 
$\hat{\y} = \frac{1}{L}\sum_{l=1}^L
g^{-1}(\hat{\alpha}_l 
+  \X_{l} \hat{\bbeta}_ l)$, where $g^{-1}()$ is the inverse link
function associated with the 
exponential family distribution.   Thus, our overall prediction is 
an ensemble of local predictions.  Note if our local models were
estimated completely separately as with $\lambda_{agg} = 0$, then this
ensemble would be the naive raw average or majority vote of local
models.  On the other hand, by using our aggregating penalty and
estimating the local models together, the local
coefficients that determine the local predictions depend on each other; hence, the coefficients of locations less useful for prediction are shrunk towards zero.  Overall, the predictions
of our Local-Aggregate model behave like an ensemble.

 \subsection{Advantages of Local-Aggregate Modeling Framework}
\label{sec:advantage}

Our method offers several
advantages in terms of statistical efficiency ({\bf S}), modeling
flexibility ({\bf M}), and computational efficiency ({\bf C}). Here,
we discuss some of these advantages as to existing
literature.

\noindent{\bf (S.1) Predictive power of ensembles.}
 The idea of ensemble learning is simple: combine
 strengths of multiple base models or weaker learners to build a set of models with
 superior predictive accuracy.  Examples of such learning
 techniques include Bayesian model
 averaging, model stacking, bagging, and
 boosting \citep{hastie2009elements}.   
 Most of these existing ensemble learning methods 
 use sampling or iterative reweighting schemes to create a diverse set
 of base learners for the same data set.  Our ensemble of local models,
 on the other hand, uses separate local data, $\X_l$, for each base
 model and hence there is no need for computationally-intensive
 sampling schemes.  
Meanwhile, our ensemble incorporates external spatial information through spatially smoothed $\hat{\bbeta}_l$, resulting in a combination of weaker yet more informative local
 learners that can potentially yield better prediction than
 one sophisticated multivariate model.

\noindent{\bf (S.2) Lower-dimensional problem.}
The matrix GLM problem is
ultra-high-dimensional with small $n$ relative to $L \tau$ parameters
as compared to $\tau$ 
parameters for each of our local models.  Thus, we can achieve much
improved statistical efficiency to estimate the parameters of our local
models than for the true matrix GLM.  
Often obtaining better estimates
of incorrect models leads to better predictive accuracy than poorer
estimates of correct models.
For example, it is well known that biased regularized linear models have better prediction accuracy under much weaker conditions than the
correct, unbiased linear model \citep{hoerl1970ridge}.  Since our Local-Aggregate model
seamlessly translates an ultra-high-dimensional problem to a
lower-dimensional one, we expect that the improved statistical
efficiency will result in better predictive performance. 
Our method is also highly scalable: as the number of
locations $L$ increases, the complexity of our local models remains constant.

\noindent{\bf (M.1) Spatio-temporal smoothness.}
Our Local-Aggregate modeling framework directly accounts for the spatio-temporal structure of the data by encouraging both spatial and temporal coefficient smoothness, thus yielding more interpretable results.
This is similar to the two-way smoothing penalizes used in the context of dimension reduction by 
Huang et al. (2009); Allen et al. (2014) and MEG signal reconstruction by \citet{tian2012two}.

\noindent{\bf (M.2) Location selection.}
Our modeling framework can select brain locations important for
prediction in a data-driven manner through the group lasso penalty;
this shrinks the coefficients at non-informative locations towards
zero and their local predictions to $\hat{\alpha}_l$.  Therefore the
aggregating penalty combined with the group lasso penalty
automatically weight the local coefficient vectors by their predictive
power, and our overall prediction is unaffected by summing over
uninformative locations. 
When considering the advantages of our method, one may question the
relative roles of modeling flexibility as compared to statistical efficiency.
To address this directly, we study our ensemble of local models and the full matrix GLM with the same local and aggregating
penalties in the Supplementary Material Section C.  Briefly, our
investigation suggests that the improvements in predictive
performance we observe are due to both the statistical and modeling advantages of
our approach.

\noindent{\bf (C) Parallelizable Algorithm.}
 Distributed storage and computation is crucial to big data modeling,
 especially when the data is too large to fit into one computer's
 memory.  However, few statistical algorithms are able to distribute
 data storage and memory at the same time.  Our Local-Aggregate
 modeling framework is intrinsically primed for parallel computing,
 discussed subsequently, as the only non-separable term is the aggregating penalty.

\section{\bf Algorithm}
\label{sec:algorithm}

Our primary goal is to build a scalable modeling and algorithmic strategy for structured big data with many locations, more specifically, large-scale regression problems with matrix-valued predictors.
Notice that as our objective function \eqref{loc_agg_gen} is jointly convex in $\B$ and $\balpha$, there are many potential optimization strategies that enable parallel computing; however, many of these do not allow for distributed memory and data storage. We desire a highly-parallelizable algorithm that permits distributed data storage and memory with minimal message passing between the computing nodes.  
To achieve this, we turn to the Alternating
 Direction Method of Multipliers (ADMM) optimization strategy.  
\subsection{Local-Aggregate ADMM }
To develop our Local-Aggregate algorithm, we begin with no local intercept $\alpha_l$ to match the classical ADMM framework outlined in \citet{boyd2011distributed}:
{\begin{equation}
\label{loc_agg_reparam}
\underset{\B}{\text{minimize}}
\quad  \sum_{l = 1}^L\big[ \ell (\y; 
      \X_{l} \bbeta_{l})+
    \lambda_{loc} P_{loc}(\bbeta_{l})\big] 
+ \lambda_{agg} P_{agg}(\B,\G).
\end{equation}
}We will show how to deal with local intercepts in detail later in Section~\ref{sec:loc_agg_neuro}.

In order to apply  ADMM to our Local-Aggregate modeling framework, we first create a copy $\Z \in \Re^{\tau \times L}$ to substitute the modeling coefficient matrix $\B$ in the aggregating penalty so that we can split the original problem \eqref{loc_agg_reparam} into the local models and the aggregating term.  Then we require an additional constraint $\B = \Z$ to ensure the two copies are the same:
{\begin{equation}
\label{loc_agg_gen2}
\underset{\B, \Z}{\text{minimize}}
\quad  \sum_{l = 1}^L\big[ \ell (\y;\X_{l} \bbeta_{l})+
    \lambda_{loc} P_{loc}( \bbeta_{l})\big] 
+ \lambda_{agg} P_{agg}( \Z,\G) \quad \text{subject to }   \B = \Z
\end{equation}
}The subproblems are solved by minimizing the {\em augmented Lagrangian} of \eqref{loc_agg_gen2} with respect to $\B$ and its copy $\Z$ independently:
{\begin{equation}
\label{lagrangian2}
\mathcal{L}_{\rho}(\B, \Z, \U) = \sum_{l = 1}^{L}\big[ \ell(\y;
       \X_{l} \bbeta_{l} ) + \lambda_{loc} P_{loc}( \bbeta_{l}) \big] 
      +\lambda_{agg}P_{agg}(\Z,\G) + \sum_{l = 1}^{L} \frac{\rho}{2}\| \bbeta_l-\z_l +\uu_l \|_2^2,
\end{equation}
}where $\U \in \Re^{\tau \times L}$ is the dual variable, and $\rho > 0$ is an algorithm parameter.  The solutions to the subproblems are then coordinated via a dual update procedure to find the  global solution.
Thus, the three key steps of our Local-Aggregate ADMM are as follows:
\newline
1. $ \bbeta_{l}^{k+1} =  \argmin_{\bbeta_l} \ell(\y; \X_{l} \bbeta_{l}) +  \lambda_{loc} P_{loc}(\bbeta_{l}) + \frac{\rho}{2}\|\bbeta_{l} - \z_{l}^k + \uu_{l}^k\|_2^2$ $\rightarrow \B\text{-subproblem (in parallel)}$
\newline
2. $  \Z^{k+1} =  \argmin_{\Z}  \lambda_{agg}P_{agg}(\Z,\G)+  \frac{\rho}{2}\sum_{l=1}^L \|\z_{l} - \bbeta_{l}^{k+1}-\uu_{l}^k\|_2^2 \rightarrow \Z \text{-subproblem (message passing)}$
3. $ \uu^{k+1}_{l} = \uu^k_{l} + \bbeta^{k+1}_{l} - \z^{k+1}_{l} \rightarrow \text{dual update (in parallel)}$
\newline \noindent We are fitting regularized GLMs to each location, for which many well-studied optimization algorithms
exist \citep{friedman2010regularization}.  If the local penalty is smooth, such as with a temporal roughness penalty, we can apply a fast Newton algorithm to solve the $\B$-subproblem, which achieves at least quadratic convergence rate.  If the local penalty is non-smooth, we then employ the proximal gradient method, which guarantees at least $\mathcal{O}(1/k)$ convergence rate under the assumption of convex loss functions (at least one being strongly convex) \citep{nesterov2007gradient}.  
Once the local regression information from previous $\B$ updates is gathered, the aggregating penalty enforces smoothness
 of coefficients over neighboring locations.
The $\Z$-subproblem collects the local coefficients and blends structural information across nearby locations through the spatial smoothness penalty, $P_{agg}(\B,\G)$. 
The smoother structural information is then distributed to all locations for the next dual update step.  The dual variable $\U$ tracks the progress of the algorithm as $\B$ is squeezed closer to $\Z$ until
the equality constraint is satisfied.  The dual update step plays an important role of coordinating $\B$ and $\Z$ towards the global solution: as the algorithm converges, $\Z$ approaches $\B$.  

Note that as the augmented Lagrangian \eqref{lagrangian2} is separable in locations, we can solve the ensemble of separate local models ($\bbeta_l$-subproblems) completely in parallel.   Another major computational advantage of our algorithm is its distributed data storage and minimal message passing among computing nodes.  To start with, we send the response $\y$ and the local covariates $\X_l$ to each computing node $l$ to compute the $\bbeta_l$ updates, which will be collected to solve the $\Z$ subproblem, then we only need to pass back the new $\z_l$'s, as all other variables are local.  Finally, the dual variable update step is local and requires no message passing.

\subsubsection{Convergence Analaysis}
Our Local-Aggregate ADMM is guaranteed to converge:
\newline
{\bf Theorem 3.1.} {\it(Extension of a result from Section 3.2 in \citet{boyd2011distributed})
Under the following assumptions,
{\em (a)} $\ell():  \Re^{\tau \times L} \rightarrow \Re$,  $P_{loc}(): \Re^{\tau} \rightarrow \Re^{+}$ and $P_{agg}():  \Re^{\tau \times L} \rightarrow \Re \cup \{+ \infty \}$ 
are closed, proper and convex,
{\em (b)} $\rho,\lambda_{loc},\lambda_{agg} \in \Re^+$ fixed, the ADMM iterates satisfy the following 
convergence:
1) Residual convergence: $\B^k - \Z^k \rightarrow 0 \text{ as } k \rightarrow \infty$.
2) Objective convergence: $\sum_{l = 1}^L[ \ell(\y; \X_{l} \bbeta_{l}^k)+
    \lambda_{loc} P_{loc}(\bbeta_{l}^k)] 
+ \lambda_{agg} P_{agg}(\Z^k,\G) \rightarrow p^*$, the optimal value. 
3) Dual convergence: $\U^k \rightarrow \U^*\text{ as }k  \rightarrow \infty$, where $\U^*$ is a dual optimal point scaled by $\rho$.
}
\newline
The convergence of the algorithm hinges upon the convexity of the GLM loss function and the penalty functions.  As all GLMs with canonical link functions are convex, and many commonly used penalty functions, e.g. lasso, ridge, elastic net, group lasso, are all convex functions, we have a very flexible choice of local GLMs and local penalty functions, and hence our Local-Aggregate ADMM algorithm is suitable for many applied statistical problems.  

The convergence rate of our Local-Aggregate ADMM is $\mathcal{O}(1/k)$, which is the same as that of ADMM \citep{deng2012global}, which can be further accelerated by using a variable penalty parameter $\rho$ depending on the algorithm progress (how fast $\B$ is pushed towards $\Z$) \citep{boyd2011distributed}.  We also develop a self-adaptive vectorized penalty parameter updating scheme which takes into account the difference in convergence speeds among all time points and automatically selects appropriate $\rho$ for each time point. This further accelerates the algorithm and is described in detail in the supplementary material.  

\subsection{Local-Aggregate ADMM Example: Multi-Subject Neuroimaging Data}
\label{sec:loc_agg_neuro}
 To illustrate the details of our algorithm framework, we study our neuroimaging inspired problem \eqref{loss_neuro}, for binary classification with local logistic loss functions. (This is also the model we employ for an EEG classification example in Section~\ref{sec:eeg}).
Let  $\x_{il} = \Xcal_{(i,:,l)} \in \Re^{\tau}$ denote measurements for subject $i$ taken at location $l$, and $\y \in \Re^n$ denote subject-level responses, we then have the following neuroimaging classification problem:
\begin{equation}
\label{loc_agg_eeg_orig}
\begin{aligned}
& \underset{\balpha, \B, \Z}{\text{minimize}} \sum_{l = 1}^{L}[  \sum_{i = 1}^n\{-\y_i(\alpha_l +\x_{il}^T \bbeta_l) + \log(1 + e^{(\alpha_l + \x_{il}^T\bbeta_l)}) \}+ \lambda_{loc}^{sm}
     \bbeta_{l}^{T} \Omeg \bbeta_{l} +  \lambda_{loc}^{sp} \|
      \bbeta_{l} \|_{2}] \\ &
    \hspace*{.3in} + \lambda_{agg} \text{ tr}(\Z \G \Z^T), \quad  \text{subject to } \B = \Z.
\end{aligned}
\end{equation}

However, as the classical ADMM algorithm only allows for two coupled sets of variables 
\citep{chen2013direct}, we need to transform the above problem into the Local-Aggregate ADMM  framework with no local intercepts through reparametrization.  Let $\widetilde{\x}_{il} = [1;\x_{il}] \in \Re^{(\tau+1)}$, $\widetilde{\bbeta}_l= [\alpha_l; \bbeta_l]  \in \Re^{(\tau+1)}$, {\small $\Sb = {\small 
\begin{pmatrix}
  0 &  \bf{0}\\
  \bf{0} &  \I_{\tau}
 \end{pmatrix}}$, $\widetilde{\Omeg} {\small = 
 \begin{pmatrix}
  0 &   \bf{0} \\
  \bf{0} &  \Omeg 
 \end{pmatrix}}$}, both $\Re^{(\tau+1)\times(\tau+1)}$ matrices.  Let $\text{Prox}_g(x,t)$
 \newline\noindent$ = I_{\{\|x\|_2 \geq t\}}(1 - t/\|x\|_2)x$ denote the proximal operator of $\ell_2$ norm \citep{yuan2006model}, let $ f(\bbeta_l)$ denote the logistic loss plus the the local smoothness penalty, and let $\rr = \bbeta_l - \z_l$ denote the primal residual, and $\s = \z^{k} - \z^{k-1}$ denote the dual residual.  Then, ignoring the $\  \widetilde{}\  $ sign for notational convenience, the Local-Aggregate ADMM for solving this classification problem \eqref{loc_agg_eeg_orig} is outlined by Algorithm 1.  (We formally prove the equivalence of this reparametrized framework to the original problem and its convergence rate in the supplementary material.)
 { 
{ 
\begin{algorithm}
\label{alg1}
\caption{\bf : Local-Aggregate ADMM for Neuroimaging Example}
\begin{small}
{\raggedright{\bf Initialize} $\B^0 = \mathbf{0}$, $\U^0 = \mathbf{0}, \Z^0 = \mathbf{0}, \gamma \in (0,1)$, $k = 1$, $\rho$ initialized.\\
{\bf While} $\rr > \epsilon^{tol}$ and $\s >\epsilon^{tol}$\\
\hspace*{.2in}{\bf For} $l = 1:L$\hspace*{2.6in} {\color{blue} $\rightarrow$ Parallel}\\
\hspace*{.4in}{\bf While} $\| \bbeta_l^k - \bbeta_l^{k-1} \| >  \epsilon^{tol} $ \\
\hspace*{.6in}$ t = 1$\\
\hspace*{.6in}$ \nabla f^{k}=\X_l^T(\y-1/e^{-\X_l\bbeta_l^k}) +2\lambda_{loc}^{sm}\Omeg\bbeta_{l}^k+  \rho \Sb( \bbeta_l^k - \z_l^{k} + \uu_l^{k})$\\
\hspace*{.6in} $\bbeta_l^{k+1} = \text{Prox}_g( \bbeta_l^{k} - t  \nabla f^{k},\lambda_{loc}^{sp} t)$\\
\hspace*{.6in} {\bf While} $f(\bbeta_l^{k} - t \nabla f^{k} ) > f(\bbeta_l^{k} ) - t ({ \nabla f^{k}})^{T}\text{Prox}_g( \bbeta_l^{k} - t  \nabla f^{k},\lambda_{loc}^{sp} t)+$  \\
 \hspace*{.8in}$\frac{t}{2}\| \text{Prox}_g( \bbeta_l^{k} - t  \nabla f^{k},\lambda_{loc}^{sp} t)\|_2^2$\\
\hspace*{1in} $t := \gamma t$\\
\hspace*{1in}$ \bbeta_l^{k+1} =\text{Prox}_g( \bbeta_l^{k} - t  \nabla f^{k},\lambda_{loc}^{sp} t)$\\
\hspace*{.6in}{\bf End while}\\
\hspace*{.4in}{\bf End while}\\
\hspace*{.2in}{\bf  End for}\hspace*{2.9in}{\color{blue} $\rightarrow$ End parallel}\\
\hspace*{.2in}$\Z^{k+1} = \rho(\Sb\B^{k+1}+\U^{k})(\rho \I_L + 2\lambda_{agg}G)^{-1}$\hspace*{.65in}{\color{red} $\rightarrow$ Message passing}\\
\hspace*{.2in}{\bf For} $l = 1:L$\hspace*{2.6in}{\color{blue} $\rightarrow$ Parallel}\\
\hspace*{.4in}$\uu_l^{k+1} =\uu_l^{k} + \bbeta_l^{k+1} -\z_l^{k+1}$\\
\hspace*{.2in}{\bf  End for}\hspace*{2.9in}{\color{blue} $\rightarrow$ End parallel}\\
\hspace*{.2in}$k = k+1$\\
{\bf End while}\\
}
\end{small}
\end{algorithm} }}
Overall, we have developed a highly-parallelizable ADMM algorithm with distributed
memory and data storage, minimal message passing, and fast convergence speed, all of which 
are well suited for our Local-Aggregate optimization framework for multi-subject neuroimaging data with many locations.

\section{\bf Model Selection: Local-Aggregate Algorithm Path}
\label{sec:alg_path}

We have carefully developed a fast, completely parallelizable algorithmic strategy for large-
 scale matrix-covariate regression problems. In practice, however, often the most computationally intensive part of fitting statistical machine learning models is not fitting at a single regularization parameter, $\lambda$, but instead performing model selection by considering a sequence of $\lambda$'s.  Typically, one considers a grid of $\blambda = \{ \lambda = 0, \ldots, \lambda_{max} \}$ and chooses a value of $\lambda$ that minimizes a selection criterion such as BIC, AIC, GCV, or employs sampling schemes such as cross-validation (CV) or stability selection at each $\lambda$.  For our problem, repeatedly fitting 50-100 models (or more for re-sampling) is computationally infeasible.  Thus, we consider a completely novel yet effective approach to model selection.

\begin{sidewaysfigure}[htbp] 
   \centering
   \includegraphics[width=8in]{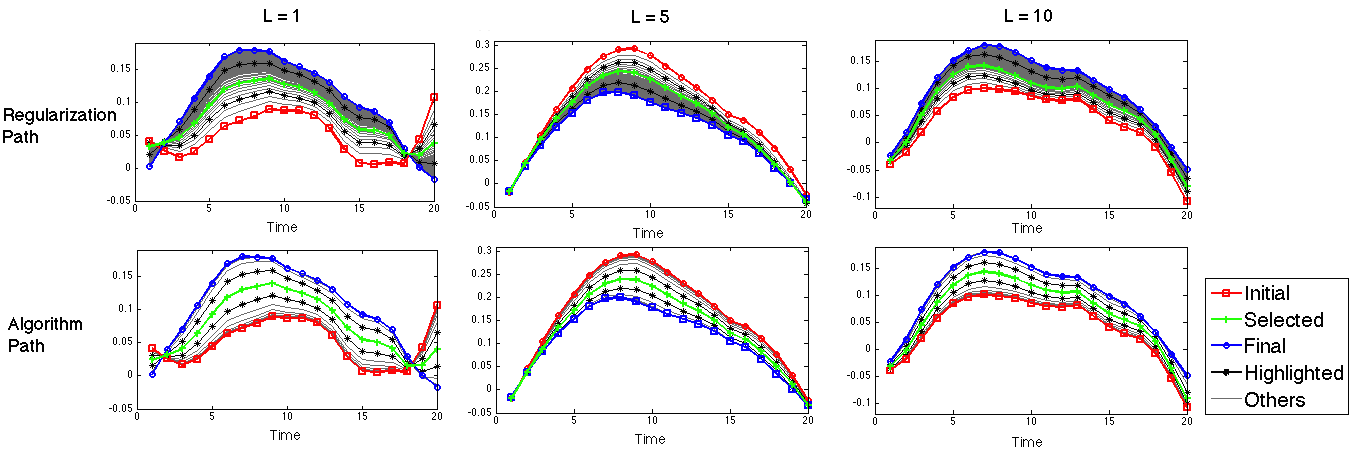} 
   \caption{ Comparison of the entire regularization path with the algorithm path at L = 1 (left panels), L = 5 (middle panels), and L = 10 (right panels). 
   ``Initial'' refers to the model at $\lambda = 0$ or $k = 1$, ``Selected'' refers to the optimal model selected via cross validation, ``Final'' refers to the model at $\lambda = \lambda_{max}$ or $k  = K$, and ``Highlighted'' refers to some highlighted model along the path for illustrative purposes, and ``Others'' refers to the rest of the path of solutions.
   This empirically demonstrates that our algorithm path well approximates the regularization path.  Also the model selected by CV is comparable for both methods.}
   \label{paths_2}
\end{sidewaysfigure}

First, for our Local-Aggregate modeling framework, notice that we can
select any $\lambda_{loc}$ completely separately for each location in
parallel.  Thus we are left with how to choose the aggregating
penalty, $\lambda_{agg}$, which cannot be done in a parallel manner.
Our goal is to develop a computationally efficient method to select
the correct amount of smoothing over locations.  Let us start with a
simulated linear regression example: suppose we have a chain graph of
$L = 10$ locations, $n = 100$ subjects, and $\tau = 20$ time
points. The classical regularization path is computed by varying
$\lambda_{agg}$, and the entire path of solutions is shown
at $L = 1,5$, and 10 in top panel of Figure~\ref{paths_2}. The
regularization path of our Local-Aggregate modeling framework
$\{\hat{\B}(\lambda_{agg}): 0 < \lambda_{agg} < \infty\}$ starts from
$\lambda_{agg} = 0$, i.e. completely independent local GLMs, to
$\lambda_{agg} =\lambda_{max}$, some large value giving the extreme
smoothness over locations. The parameter $\lambda_{agg}$ controls the
amount of smoothing over locations: as $\lambda_{agg}$ increases,
there is increased smoothness over locations.  

If we take a closer look at the iterates of our Local-Aggregate ADMM
algorithm when we set $\lambda_{agg} = \lambda_{max}$ plotted in
the bottom panel of Figure~\ref{paths_2}, we see that the iterates of
our algorithm look very similar to the regularization path for
different $\lambda$ values.   
In particular, for almost all $\hat{\B}(k)$ (our estimate from the
$\B$-subproblem at iterate 
$k$ of the ADMM algorithm), there exits some corresponding
$\hat{\B}(\lambda_j)$ in the regularization path that precisely
matches $\hat{\B}(k)$. 
This observation makes sense considering how our Local-Aggregate ADMM
algorithm progresses.  Namely, at iteration 1, we start with $\Z
= \0$, giving no smoothness and a fully local solution for $\B$.  Thus
at iteration 1, $\hat{\B}(k)$ is equal to the regularization path for
$\hat{\B}(\lambda_j = 0)$.  As the ADMM algorithm progresses,
$\hat{\B}(k)$ gets squeezed towards the overly smoothed $\Z$, meaning
that at each subsequent iteration, more smoothness is induced.  This
occurs until finally the ADMM algorithm converges to the solution at
$\lambda_{agg} = \lambda_{max}$, meaning that $\hat{\B}(k =
K^{\text{final}}) = \hat{\B}(\lambda_j = \lambda_{max})$.  While we
have observed this correspondence between our ADMM iterates and the
regularization path over a grid of $\lambda$'s in the simulation in
Figure~\ref{paths_2} as well as all other empirical investigations,
proving such equivalence is beyond the score of this paper; we hope to
investigate this formally in future work.

Hence based on our empirical observations, we propose the so-called Local-Aggregate Algorithm Path, which approximates the solutions to the regularization path over a range of $\lambda$'s with the iterates of our Local-Aggregate ADMM algorithm with fixed tuning parameter at $\lambda_{agg} = \lambda_{max}$.  The difference between the regularization path and the local-aggregate algorithm path is illustrated as follows:
\newline
 {\bf  $\quad \bullet$ Regularization Path}
Given a sequence of regularization parameters $ \lambda_N = \lambda_{max} > \cdots > \lambda_j> \cdots > \lambda_1 = 0$, we solve a sequence of minimization problems corresponding to each $\lambda_k$ and obtain $\{\hat{\B}(\lambda_j): j = 1, \cdots, N\}$.
\newline
 {\bf $\quad \bullet$ Algorithm Path} Given the fixed regularization parameter $\lambda_{agg} = \lambda_{max}$, we solve one minimization problem and take the series of algorithm iterates as our solution path: $\{\hat{\B}(k): k = 1, \cdots, K\}$, where $K$ is the number of iterations.

 While the grid search method evaluates the optimization problem over a sequence of $\lambda_j$'s, our local-aggregate algorithm path approximates the solutions as the algorithm iterates without tuning the regularization
parameter $\lambda_{agg}$. 
The computational cost is thus greatly reduced: namely, from solving $N$ optimization problems to solving just one problem.

\subsection{Model Selection via Local-Aggregate Algorithm Path}

The computational cost is huge if we employ M-fold 
cross-validation along with the regularization path to select the optimal model: we need to solve a sequence of $N$ minimization problems for all M cross-validated folds, totaling $M\times N$ model fits.  However, if we combine the idea of the Local-Aggregate algorithm path with cross-validation (Algorithm 2), the number of model fits is cut down to only M.
\begin{algorithm}
\caption{Compute Local-Aggregate Algorithm Path via Cross Validation}
\begin{enumerate}
\item[1.] Set $\lambda_{agg}$ large, split data into M folds
\item[2.] For $m = 1, \cdots, M$, run local-aggregate algorithm on subsets of data with the $m^{th}$ fold left out.  Record $\hat{\B}(k)$ for each iteration of the algorithm.
\item[3.] Compute CV error for $\hat{\B}(k)$ generated by each iterate on the $m^{th}$ left out fold.
\item[4.] Average the CV errors over M folds, and obtain the CV error curve across iterates of the algorithm.
\item[5.] Pick the optimal iteration $K^{opt}$ that has the minimum CV error.
\end{enumerate}
\end{algorithm}
Unlike traditional model selection methods, which seek to choose a $\lambda^*$, our proposed model selection strategy chooses the correct model from one of the algorithm iterates.  
While we have not formally proven this, we conjecture that our algorithm iterates correspond to the solution at a specific $\lambda$ value, and thus our model selection procedure is equivalent to M-fold CV.  
Overall, our algorithm path and model selection procedure represent a completely novel approach that offer substantial computational savings.

\section{\bf Simulation}
\label{sec:sim}

To better understand the performance of our Local-Aggregate modeling framework, and how it compares with the classical methods, we present a series of simulated examples inspired by neuroimaging data. The true spatio-temporal signal $\B_o = \sum_{r = 1}^R \vv_r \uu_r^T$ is simulated by the sum of outer products of a spatial factor, $\uu_r \in  \Re^{L}$ corresponding to a  $\sqrt{L} \times \sqrt{L}$  grid consisting of zeros except for two blocks (one near either end of the grid's diagonal),
 and  a temporal factor $\vv_r \in \Re^{200}$ with 200 equally-spaced time points following sinusoidal curves, for $r = 1, \dots, R$, where $R = \text{rank}(\B_o)$.  Note that $\{\vv_1, \cdots, \vv_R\}$ and $\{\uu_1, \cdots, \uu_R\}$ are sets of orthogonal vectors so that the rank of the true signal is $R$. The tensor-valued covariates $\Xcal$, are a collection of covariates $\X_i \overset{iid}{\sim} N(0,\Sigma_L \otimes \Sigma_T)$ for each subject, $i$, generated as matrix-variate normal with exponential spatial ($\Sigma_L = e^{-\D_L^2/\theta_L}$) and temporal ($\Sigma_T = e^{-\D_T^2/\theta_T}$) covariance structures.  Here, $\D_L$ and $\D_T$ are spatial and temporal distances respectively, and $\theta_L$ and $\theta_T$ control the amount of spatial and temporal correlation.  
\subsection{Regression simulation}
For our regression simulation, the response variable is generated as $\y =\X_{(1)} vec(\B_o)+\epsilon$, for $\epsilon \overset{iid}{\sim} N(0,1)$.  
We compare our method to: the Lasso, Ridge, Elastic Net, and nuclear-norm regularized tensor regression method \citep{zhou2013tensor}. Since the classical regression methods take vectors as covariates, the time series by locations are vectorized. We demonstrate our findings in three examples. First, we simulate data with differing numbers of locations, keeping the rank of the spatio-temporal signal the same.  Second, we compare methods for different ranks of $\B_o$ with fixed numbers of locations, $L = 100$. Third, we explore the effects of the amount of spatio-temporal correlation in the covariates for the case of $L = 100$, and $\text{rank}(\B_o) = 2$. 

We present simulation results for fifty replicates and compare various methods in terms of both prediction accuracy and signal recovery. We use mean squared error (MSE) as the criterion for prediction error, and to evaluate signal recovery, we use MSE of the estimated
coefficient matrix, $\| \hat{\B} - \B_o\|_2$, and the true/false positive rates of the true location detection. 
\begin{table}
\caption{ \small Linear regression results for varying (top) number of locations, with $n = 100$, $\tau = 200$, $rank(\B_o) = 2$, $SNR = 10$; (middle) rank of signal, with $n = 100$, $\tau = 200$, $L = 100$, $SNR = 10$; and (bottom) spatio-temporal correlation, with $n = 100$, $\tau = 200$, $L = 100$, $rank(\B_o) = 2$, $SNR = 10$.}
\label{table:table1}
\begin{center}
{\scriptsize\begin{tabular}{ l  l  l*{4}l }
\hline
&   &  Method & Prediction Error  & MSE(B) & FPR & TPR\\
\hline
\hline
L& 25 &Loc-Agg &0.4857(0.0208)&0.7932(0.0174)&47.52\%(4.22\%)&94\%(2.09\%)\\
  && Lasso   &0.6978(0.0307)&1.1896(0.0095)&47.33\%(3.10\%) &88.50\%(3.99\%)\\
&&Ridge &0.5789(0.0161)&0.7761(0.0098)& / & /  \\
&&Elastic Net    &0.5448(0.0218)&0.9871(0.0082)&81.71\%(1.31\%)&100.00\%(0\%)\\
 \rowcolor{Tgray}&&Tensor Reg.   &0.3913(0.0200)&0.8645(0.0165)& / & / \\
\hline
 \rowcolor{Tgray}& 64 &Loc-Agg &0.5290(0.0217)&0.7891(0.0180)&29.29\%(2.85\%)&92.25\%(2.11\%)\\
  && Lasso   &0.8289(0.0274)&1.1424(0.0064)&23.07\%(1.91\%) &69.25\%(4.24\%)\\
&&Ridge &0.7026(0.0193)&0.8358(0.0108)& / & /  \\
&&Elastic Net    &0.7340(0.0252)&1.0441(0.0099)&51.86\%(1.00\%)&96.75\%(0.78\%)\\
&&Tensor Reg.   &0.5554(0.0270)&0.9358(0.0179)& / & / \\
\hline
 \rowcolor{Tgray}  &100 &Loc-Agg &0.5209(0.0275)&0.8071(0.0177)&19.98\%(3.67\%)&88.50\%(2.47\%)\\
  && Lasso   &0.8989(0.0300)&1.1151(0.0087)&12.76\%(1.24\%) &56.25\%(4.52\%)\\
&&Ridge &0.7768(0.0212)&0.8995(0.0115)& / & /  \\
&&Elastic Net    &0.7807(0.0253)&1.0702(0.0092)&38.52\%(0.78\%)&92.50\%(1.43\%)\\
&&Tensor Reg.   &0.6952(0.0254)&1.0165(0.0107)& / & / \\
\hline
   \rowcolor{Tgray}&144 &Loc-Agg &0.5889(0.0284)&0.8409(0.0182)&16.56\%(2.78\%)&89.00\%(2.52\%)\\
  && Lasso   &0.9438(0.0312)&1.1282(0.0089)&9.01\%(0.99\%) &49.25\%(4.19\%)\\
&&Ridge &0.8597(0.0264)&0.9396(0.0108)& / & /  \\
&&Elastic Net    &0.8461(0.0293)&1.0939(0.0088)&30.97\%(0.60\%)&87.25\%(1.62\%)\\
&&Tensor Reg.   &0.8193(0.0296)&1.0807(0.0123)& / & / \\
\hline\hline
  \rowcolor{Tgray} Rank($\B_o$)&  2 &Loc-Agg &0.5209(0.0275)&0.8071(0.0177)&19.98\%(3.67\%)&88.50\%(2.47\%)\\
 && Lasso   &0.8989(0.0300)&1.1151(0.0087)&12.76\%(1.24\%) &56.25\%(4.52\%)\\
&&Ridge &0.7768(0.0212)&0.8995(0.0115)& / & /  \\
&&Elastic Net    &0.7807(0.0253)&1.0702(0.0092)&38.52\%(0.78\%)&92.50\%(1.43\%)\\
&&Tensor Reg.   &0.6952(0.0254)&1.0165(0.0107)& / & / \\
\hline
   \rowcolor{Tgray}&4 &Loc-Agg &0.5682(0.0285)&0.8424(0.0197)&15.87\%(2.48\%)&80.00\%(3.09\%)\\
  && Lasso   &0.9549(0.0270)&1.0952(0.0081)&11.22\%(1.35\%) &46.75\%(4.64\%)\\
&&Ridge &0.8009(0.0241)&0.9041(0.0139)& / & /  \\
&&Elastic Net    &0.8306(0.0277)&1.0465(0.0101)&38.74\%(0.84\%)&91.50\%(1.49\%)\\
&&Tensor Reg.   &0.7073(0.0277)&1.0104(0.0141)& / & / \\
\hline
 \rowcolor{Tgray} & 8 &Loc-Agg &0.6011(0.0251)&0.8170(0.0161)&20.17\%(2.91\%)&86.50\%(2.92\%)\\
  && Lasso   &0.9446(0.0295)&1.1075(0.0086)&10.20\%(1.43\%) &50.25\%(4.76\%)\\
&&Ridge &0.8352(0.0227)&0.9018(0.0130)& / & /  \\
&&Elastic Net    &0.8552(0.0251)&1.0612(0.0098)&37.72\%(0.86\%)&94.00\%(1.30\%)\\
&&Tensor Reg.   &0.7594(0.0238)&1.0227(0.0134)& / & / \\
\hline\hline
 Correlation&None &Loc-Agg &1.0079(0.0211)&1.0844(0.0076)&77.52\%(5.52\%)&79.00\%(5.35\%)\\
  && Lasso   &1.0420(0.0218)&1.0404(0.0073)&4.17\%(1.01\%) &4.25\%(1.27\%)\\
&&Ridge &1.0281(0.0208)&1.1199(0.0100)& / & /  \\
&&Elastic Net    &1.0549(0.0213)&1.1273(0.0073)&84.28\%(0.61\%)&86.75\%(1.53\%)\\
 \rowcolor{Tgray}&&Tensor Reg.   &1.0016(0.0204)&1.268(0.0123)& / & / \\
\hline
 \rowcolor{Tgray}  &Small &Loc-Agg &0.7402(0.0240)&0.9052(0.0121)&41.41\%(3.93\%)&89.00\%(2.87\%)\\
  && Lasso   &1.0008(0.0199)&1.0775(0.0104)&7.89\%(1.38\%) &30.75\%(4.44\%)\\
&&Ridge &0.9225(0.0200)&1.0054(0.0104)& / & /  \\
&&Elastic Net    &0.9419(0.0224)&1.1059(0.0077)&42.96\%(0.84\%)&84.00\%(1.75\%)\\
&&Tensor Reg.   &0.8925(0.0159)&1.1419(0.0090)& / & / \\
\hline
 \rowcolor{Tgray}&  Large &Loc-Agg &0.5209(0.0275)&0.8071(0.0177)&19.98\%(3.67\%)&88.50\%(2.47\%)\\
 & & Lasso   &0.8989(0.0300)&1.1151(0.0087)&12.76\%(1.24\%) &56.25\%(4.52\%)\\
&&Ridge &0.7768(0.0212)&0.8995(0.0115)& / & /  \\
&&Elastic Net    &0.7807(0.0253)&1.0702(0.0092)&38.52\%(0.78\%)&92.50\%(1.43\%)\\
&&Tensor Reg.   &0.6952(0.0254)&1.0165(0.0107)& / & / \\
\hline
\end{tabular}}
\end{center}
\end{table}

{\em 5.1.1 Number of Locations.}
Our first set of simulations varies the number of locations: $L = 25, 64, 100, 144$. The results shown in Table 1 (top) demonstrate that our Local-Aggregate modeling framework outperforms other methods when the number of locations is large. Note that as $p = L \times \tau = 28800 \gg n = 100$, this represents an ultra-high-dimensional problem, and the performance of other methods in terms of prediction error declines as $L$ increases.  The better performance of our method when $L$ is large can be easily explained: as the number of base models increases, our ensemble is able to predict the subject-level response more accurately.   Additionally, when $L$ is large, more external spatial information is brought into the modeling framework, therefore we expect more accurate predictive results.

{\em 5.1.2 Signal Complexity.}
The signal complexity is simulated as the ranks of the original signal $\B_o$.  In the rank 2 case, the spatial signal has two major areas of interest, and the corresponding temporal signals are generated by 200 equally spaced time points following the cosine curves $cos(2 \pi)$ and $cos(4\pi)$, within $[0,1)$. We use sinusoids of slightly different frequencies at the locations in the same non-zero area to generate higher rank signals. As demonstrated in Table 1 (middle), our method performs well for complex, higher-rank signals. Other regularized regression methods give declining performance for complex signals.

 {\em 5.1.3 Amount of Spatio-Temporal Correlation.}
We compare the following three cases: regression covariates with no spatio-temporal correlation, i.e., $\X_i \overset{iid}{\sim} N(0,\I \otimes \I)$, small spatio-temporal correlation ($\theta_T = 100, \theta_L = 1$), and large spatio-temporal correlation ($\theta_T = 200, \theta_L = 2$).
Table 1 (bottom) shows that when there is no correlation in the data, our method suffers in terms of feature selection, while Lasso and Elastic Net can detect some non-zero locations. Note that all methods have about the same true and false positive rates when there is no correlation in the data, which indicates that the non-zero locations are chosen at random as expected. When we have highly correlated data, our Local-Aggregate modeling framework has the best prediction error and signal recovery as we directly account for the spatio-temporal structure of the data.  

We have also conducted a similar series of classification simulations with differing numbers
 of location and signal complexities.  However, as the message of the results is similar, the 
classification results are presented in the supplementary material.

\section{\bf Case Study: EEG Data}
\label{sec:eeg}

We demonstrate the utility of our methods for modeling multi-subject
neuroimaging data 
through a case study on Electroencephalography
(EEG).  We use a well-studied data set from
http://archive.ics.uci.edu/ml/datasets/EEG+Database.  The data set consists of 122 subjects
with 77 alcoholics and 45 controls.  We study the single-stimulus EEG
recordings averaged over 120 trials while subjects were shown an
image.  The resulting data set consists of 122 subjects by 57 channels (that have known coordinates out of the 64 channels in the data)
by 256 time points forming our predictor tensor $\Xcal$.  The
objective is to use this multi-subject EEG data to predict the binary
response, $\y$, indicating the subject's alcoholic status.  We apply
our logistic Local-Aggregate model to this data as specified in \eqref{loss_neuro}.
For our local
penalties, we employ the group lasso penalty to induce sparsity in the brain
locations and a roughness penalty on the times series to
enforce temporal smoothness.  
To construct the smoothing matrix of our aggregating penalty, $\G$, we
use spherical distances between the locations of the electrodes on the
scalp. Specifically, if $D_{l,l'}$ is the polar
distance between node $l$ and $l'$, then we define the spatial kernel smoothing weights as
 $w_{l,l'} = \text{exp}^{-D_{l,l'}^2/\theta}$ (here, we take $\theta = .1$), 
 and $\G = deg(\W) - \W$, where $deg(\W)$ is the degree matrix of $\W$ with $deg(\W)_{l,l}= \sum_{l'=1}^L w_{l,l'}, deg(\W)_{l \neq l'} = 0$.  We select the $\lambda_{loc}$ locally via cross-validation, and compute the Local-Aggregate algorithm path via CV
 to select the optimal overall model.

We compare the prediction accuracy of our
method to that of other competing classification techniques using
5-fold cross-validation and repeat this procedure many times.
%
The average misclassification rates along with standard errors
are given in Table 3.  We compare our Local-Aggregate
model using logistic regression to standard classification methods 
such as the linear SVM, logistic lasso, and logistic elastic net, as well as the tensor regression
method \citep{zhou2013regularized}. Cross-validation was used to
select all tuning parameters for all methods.  Results reveal that our method performs
best in terms of classification accuracy.  We expect this better prediction accuracy of our method since it is in fact an ensemble of local models, and fully exploits the spatio-temporal structure of the EEG data.
\begin{sidewaysfigure}
   \centering
  \includegraphics[width=8in]{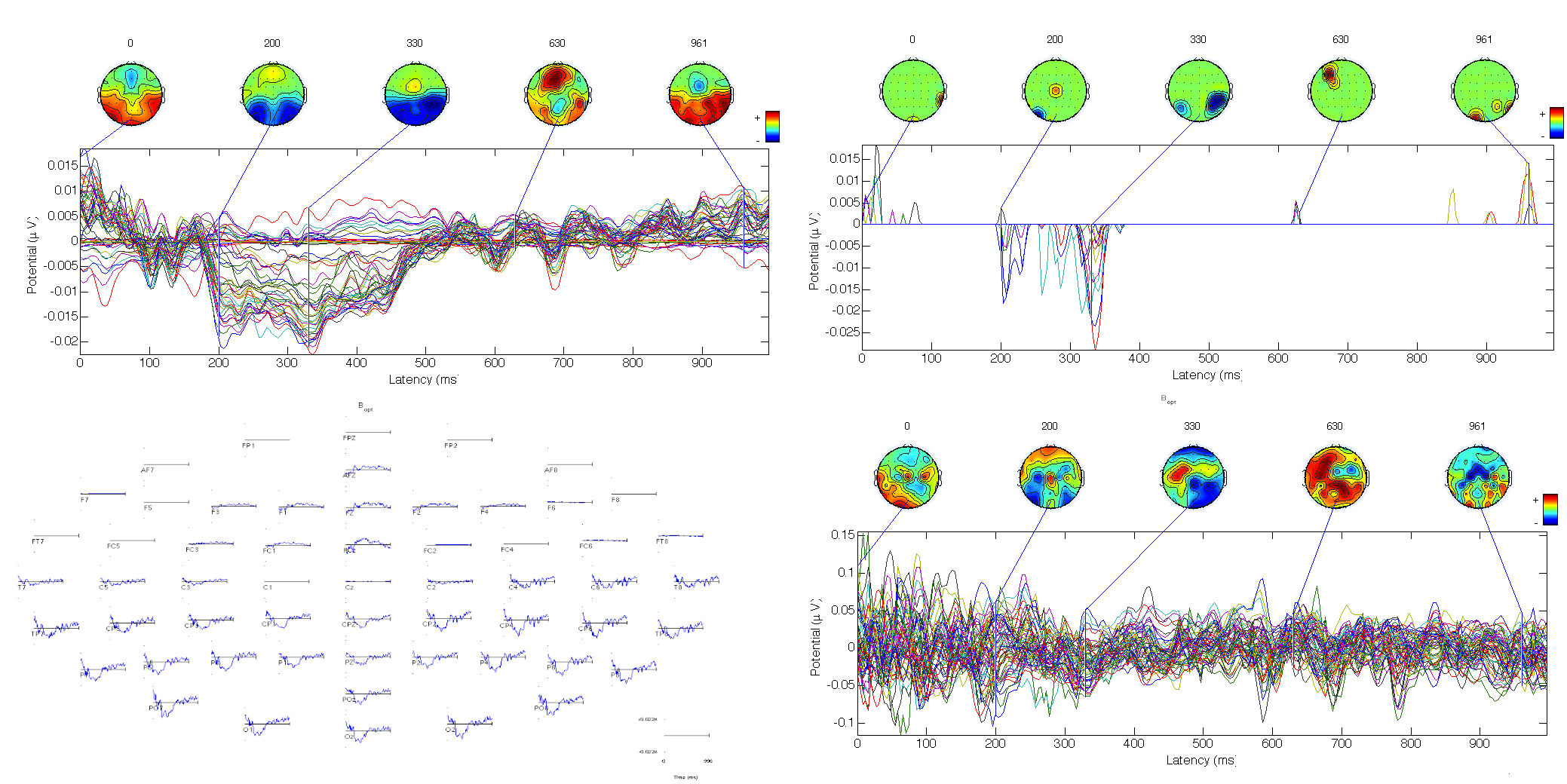} 
\caption{Case study results for the EEG alcoholism study.  Left Panel:
  Coefficients of our Local-Aggregate Model visualized by time series for all
  locations with four scalp maps shown (top) and with the time series for each
  location arrayed on a scalp map (bottom).  Equivalent time series
  with the scalp maps are shown for the elastic
  net (top right) and tensor regression (bottom right).  Our method
  yields scientifically interpretable results, showing that bi-lateral activation in the posterior brain is indicative of alcoholism}
   \label{fig_eeg}
\end{sidewaysfigure}

One major motivation of our approach is to achieve more
scientifically interpretable results.  To this end, we investigate the coefficients, $\B$,
estimated via our procedure as well as that of competitors in
Figure~\ref{fig_eeg}.  Note that $\B$ is a $\tau \times L$ matrix of
times series by locations.  Hence, we plot all 57 times series (rows of
$\B$) in the top and right portions of Figure~\ref{fig_eeg} along with five
representative scalp maps (columns of $\B$).  Notice that due to the
ultra high-dimensionality of the data, the elastic net (top right)
estimates overly sparse coefficients; on the other hand, the low-rank
constraint of tensor regression (bottom right) results in highly
variable coefficients.  Our method, however, results in interpretable coefficients, as also seen by the time series arrayed on
the scalp map (bottom left).  Several studies in addiction literature have found both latency delays and amplitude reduction in visually-evoked and event related potentials among alcoholics compared with controls \citep{holguin1999visual,porjesz2003alcoholism}.   
Our method finds that alcoholics
exhibited lower electrical activity compared to
controls in the posterior right and left hemispheres, where the visual cortex lies. Moreover, we can see delays in the activation pattern in the visual cortex among alcoholics compared with controls. Both our findings are consistent with the established literature on the effects of alcoholism on the brain activities after a visual stimulus.

Adding regularization terms to enforce smoothness over the brain locations
and time series results in less coefficient variability, and thus
more interpretable results.  Moreover, our algorithm path selects a better model with the proper amount of smoothing over brain locations.  Also notice, that by encouraging
sparsity through a group-lasso penalty, our method estimates zeros for
several portions of the anterior brain indicating that these areas do
not exhibit any difference in electrical activity between alcoholics
and controls.
\begin{table}
\label{tab_eeg}
\caption{Prediction Misclassification Error for EEG data }
\begin{center}
\begin{tabular}{  l  l }
\hline
  Method & Prediction Error  \\
\hline
  \rowcolor{Tgray}Loc-Agg &21.65\%(0.78\%)\\
   Lasso   &26.90\%(1.19\%)\\
SVM &23.93\%(1.11\%) \\
Elastic Net    &26.03\%(1.06\%)\\
Tensor Reg.   &29.32\%(1.81\%) \\
\hline
\end{tabular}
\end{center}
\end{table}

Overall, this case study has demonstrated the strengths of
our Local-Aggregate modeling framework, namely improved predictive accuracy and
scientifically more interpretable results.  

\section{\bf Discussion}

We have proposed a novel Local-Aggregate modeling framework that translates an ultra-high-dimensional tensor problem into an ensemble of local matrix problems of lower dimensions via regularization.  Our modeling framework directly accounts for the spatio-temporal structure of the data through a flexible choice of local penalties and an aggregating penalty that incorporates external spatial information into the modeling framework.  From a computational perspective, our highly parallelizable Local-Aggregate ADMM algorithm allows for both distributed memory and data storage, as well as a novel model selection strategy, the Local-Aggregate algorithm path, which chooses the correct model from one of the algorithm iterates, thus greatly reduces the computational burden of model selection.  Overall, our Local-Aggregate modeling framework and algorithmic strategy allow us to fit predictive models for large-scale multi-subject neuroimaging data in a fully distributed manner with improved prediction accuracy and scientifically more interpretable coefficients.

While we have presented an EEG application, our methods are ideally suited to predictive models for multi-subject fMRI data.  With fMRI data, there are many more locations with often hundreds of thousands of voxels.  Based on our experimental results, the larger the number of locations, the better our prediction results.  Applications to fMRI data are thus a promising area of future research.  Although we are primarily motivated by neuroimaging data, our method is applicable to other structured big data that is collected and stored in a distributed
manner, e.g. climate data (weather station), video surveillance data (images), and online shopping data (merchandise).
On the computational side, our algorithm is conducive to massive parallel computing via GPUs. Finally statistically, we have introduced a completely novel approach to model selection via the ADMM algorithm path.  While empirically we have shown that these well approximate regularization paths, much more work is needed to theoretically study these paths.

In conclusion, our work on Local-Aggregate modeling framework for modeling tensor-valued data has many implications and has opened new possibilities for research both methodologically and in application to high-dimensional tensor data and model selection.

\section*{\bf Acknowledgement}
The authors acknowledge support from NSF DMS 1209017, 1317602, and 1264058, and thank Hadley Wickham and Wotao Yin for helpful discussions, and the anonymous reviewers and AE for many helpful suggestions.

\newpage
\appendix
\renewcommand\thefigure{\thesection.\arabic{figure}}    
\renewcommand\thetable{\thesection.\arabic{table}}    
\setcounter{figure}{0}    
\setcounter{table}{0}    

{\center {\bf \large Supplementary Material for
"Local-Aggregate Modeling for \\ Big-Data via Distributed Optimization:\\ \hspace{1.8in}Applications to Neuroimaging"}}
\vspace{1in}
{\center
{\bf \small Yue Hu$^{*}$}\\
{\footnotesize Department of Statistics, Rice University}\\
{\footnotesize $^{*}$email: yue.hu@rice.edu} \\
\vspace{.5in}

	   {\bf and}\\
	   \vspace{.5in}

	   {\bf \small Genevera I. Allen$^{*}$}\\
{\footnotesize Dobelman Family Junior Chair}\\
{\footnotesize Departments of Statistics and Electrical \& Computer Engineering, Rice
University, }\\
{\footnotesize Department of Pediatrics-Neurology, Baylor College of Medicine,} \\
{\footnotesize Jan and Dan Duncan Neurological Research Institute, Texas Children's
Hospital.}\\ {\footnotesize $^{*}$email: gallen@rice.edu}

}
\newpage

\section{Supplemental Figures}
\subsection{Regression Simulation Example}
\begin{figure}[h] 
   \centering
   \includegraphics[width=5in]{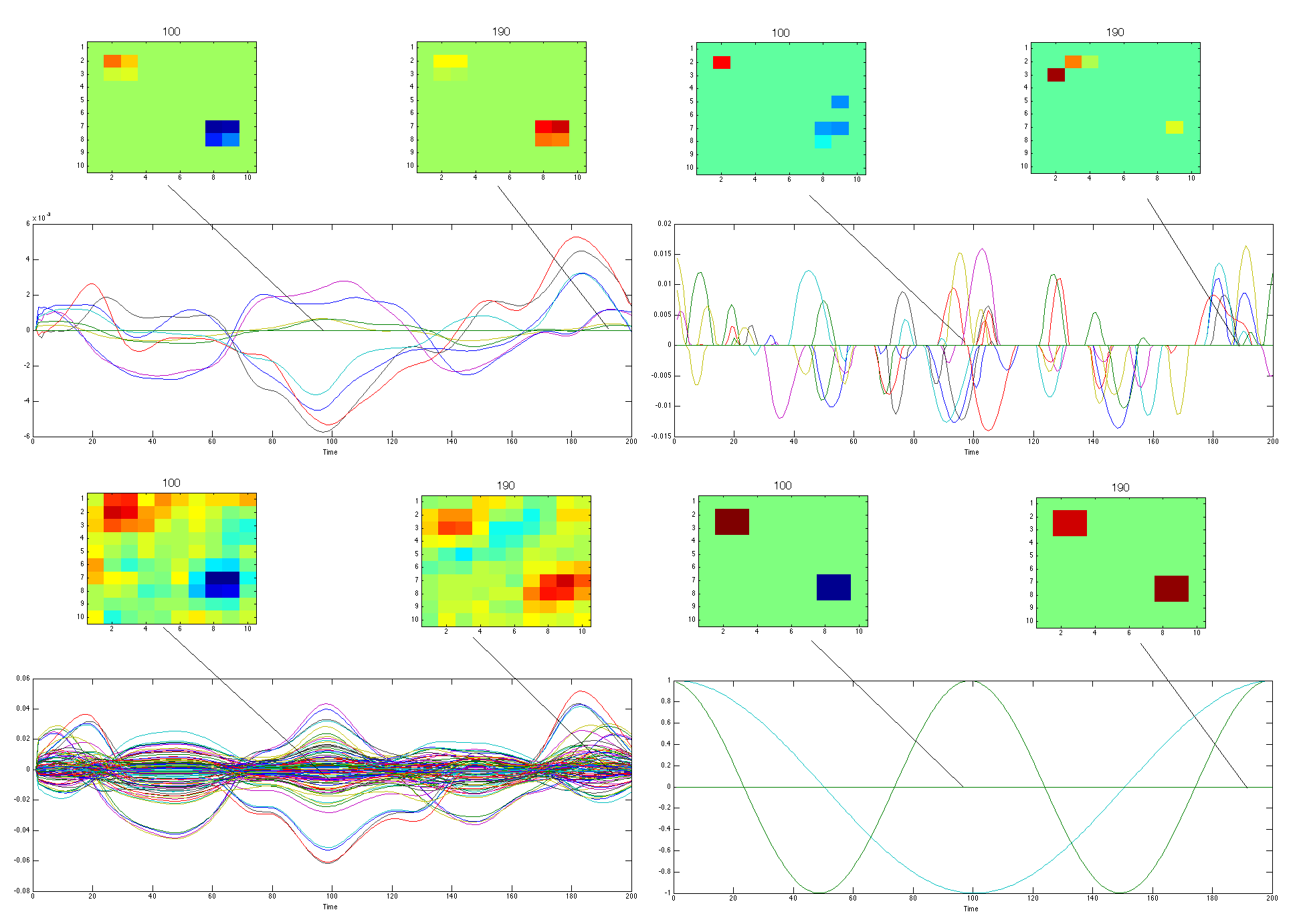} 
   \caption{Comparison of Local-Aggregate method coefficient recovery (top left) with Elastic Net (top right), Tensor Regression (bottom left) and original signal (bottom right). Each subplot shows the time signals at all locations over time with two location signals shown at time point 100 and 190. }
   \label{sim_fig}
\end{figure}

Figure~\ref{sim_fig} illustrates the coefficient matrix $\B$ from the regression simulation example in the paper.  We can see that our Local-Aggregate modeling framework correctly selects the locations and achieves locally smooth temporal coefficients at the same time.  On the other hand, lasso over-sparsifies the coefficients both in the spatial and temporal domain, regularized tensor regression cannot perform feature selection in the spatial domain.  

\subsection{Algorithm Path}
\begin{sidewaysfigure} 
   \centering
   \includegraphics[width=9in]{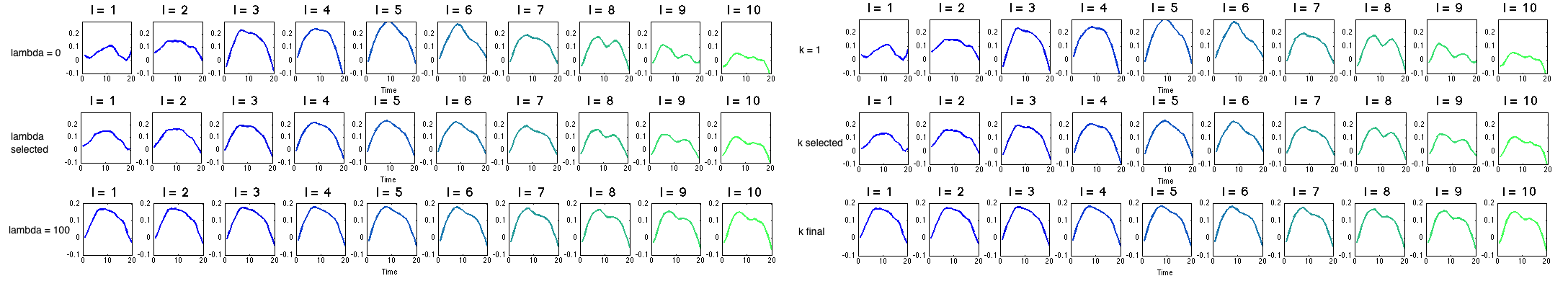} 
   \caption{Comparison of locational coefficients along regularization path (left panels) and algorithm path (right panels) at initial (up), optimal (mid), final (down) stage}
   \label{paths}
\end{sidewaysfigure}

 We can see in Figure~\ref{paths} that the initial, optimal, and final models match exactly for our algorithm path and traditional regularization path in a small simulated example. 

\section{Reparametrizing the Local-Aggregate ADMM framework}
We have the following neuroimaging classification problem:
 
\begin{equation}
\label{loc_agg_eeg_orig}
\begin{aligned}
& \underset{\balpha, \B, \Z}{\text{minimize}} \sum_{l = 1}^{L}[  \sum_{i = 1}^n\{-\y_i(\alpha_l +\x_{il}^T \bbeta_l) + \log(1 + e^{(\alpha_l + \x_{il}^T\bbeta_l)}) \}+ \\
 & \hspace*{.8in} \lambda_{loc}^{sm}
     \bbeta_{l}^{T} \Omeg \bbeta_{l} +      \lambda_{loc}^{sp} \|
      \bbeta_{l} \|_{2}] 
        + \lambda_{agg} \text{ tr}(\Z \G \Z^T) \\
   &      \text{subject to } \B = \Z\\
\end{aligned}
\end{equation}
However, as the ADMM algorithm only allows for two coupled sets of variables, we need to transform the above problem into the Local-Aggregate ADMM framework with no local intercepts through reparametrization.  
Let $\widetilde{\x}_{il} = [1;\x_{il}] \in \Re^{(\tau+1)}$, $\widetilde{\bbeta}_l = [\alpha_l; \bbeta_l]  \in \Re^{(\tau+1)}$, $\Sb = {\small 
\begin{pmatrix}
  0  &\bf{0} \\
  \bf{0} &  \I_{\tau}
 \end{pmatrix}} \in \Re^{(\tau+1)\times(\tau+1)}$, $\widetilde{\Omeg} {\small = 
 \begin{pmatrix}
  0 &   \bf{0} \\
  \bf{0} &  \Omeg 
 \end{pmatrix}} \in \Re^{(\tau+1)\times(\tau+1)}$,  we then have an ADMM framework with only 2 variables:

\begin{equation}
 \label{loc_agg_eeg}
\begin{aligned}
&\underset{\B, \Z}{\text{minimize}}\quad  \sum_{l = 1}^{L}\big[  \sum_{i = 1}^n\{-\y_i(\widetilde{\x}_{il}^T \widetilde{\bbeta}_l) + \log(1 + e^{\widetilde{\x}_{il}^T\widetilde{\bbeta}_l})  \}+  \lambda_{loc}^{sm}  \widetilde{\bbeta}_{l}^{T}\widetilde{\Omeg} \widetilde{\bbeta}_{l} +   \lambda_{loc}^{sp} \|\Sb
      \widetilde{\bbeta}_{l} \|_{2} \big]    \\ 
 &  \hspace*{.8in}    + \lambda_{agg} \text{ tr}(\Z \G \Z^T) \\
     &\text{   subject to }\Sb\widetilde{\B} = \Z.\\
\end{aligned}
\end{equation}

Ignoring the $\  \widetilde{}\  $ sign for notational convenience, the augmented Lagrangian of \eqref{loc_agg_eeg} is now given by 
\begin{equation}
\label{lagrangian_eeg}
\begin{split}
&\mathcal{L}_{\rho}(\B, \Z, \U) = \sum_{l = 1}^{L} \big[  \sum_{i = 1}^n\{-\y_i( \x_{il}^T \bbeta_l) + \log(1 + e^{\x_{il}^T\bbeta_l})  \}   + \lambda_{loc}^{sm}\bbeta_{l}^{T} \Omeg \bbeta_{l} +\lambda_{loc}^{sp} \|
     \Sb \bbeta_{l} \|_{2}  \big] \\&  \hspace*{.8in}
      +\lambda_{agg} \text{tr}(\Z\G\Z^T) + 
       \sum_{l = 1}^{L} \frac{\rho}{2}\|\Sb \bbeta_l-\z_l +\uu_l \|_2^2.
      \end{split}
\end{equation}
\newline
{\bf Lemma   B.1.} {\it Minimizing $\mathcal{L}_{\rho}(\B, \Z, \U)$ \eqref{lagrangian_eeg} is equivalent to solving the original problem \eqref{loc_agg_eeg_orig} with local intercept term $\balpha_l$.}
\vspace*{.1in}

 Recall that we need to solve the $\B$-subproblem, $\Z$-subproblem, and then update the dual variable iteratively in order to solve this ADMM optimization framework.  We employ the proximal gradient method to solve the local penalized logistic regression as there is a non-smooth local penalty $\| \bbeta_l\|_2$.  Also, the $\Z$-subproblem has an analytical solution:
\vspace*{.1in}
\newline
{\bf Lemma   B.2.} {\it  The solution to the $\Z$-subproblem is
 $\Z^{k+1} = \rho(\Sb\B^{k+1}+\U^{k})$
 $(2\lambda_{agg}G+\rho \I_L)^{-1}$.}
 \vspace*{.1in}
 \newline
\noindent Note that the matrix inverse $(2\lambda_{agg}G+\rho \I_L)^{-1}$ can be precomputed and since $\G$ is often sparse, the matrix inversion can be done quickly using fast algorithms (see \citet{li2008computing}).

\section{Local Models Vs Multivariate Model}

 We assume that the generative model for the pair of responses and tensor covariates, $(\y, \Xcal)$, follow a matrix generalized linear model (GLM) : $g(\boldsymbol{\mu}) = \alpha + \X_{(1)}^T \mathrm{vec}(\B) $, where $\boldsymbol{\mu} = \mathbb{E} (\y | \Xcal )$ is the conditional mean responses, $\alpha \in \Re$ is the intercept, and
$\B \in \Re^{\tau \times L}$ is the coefficient matrix or
$\mathrm{vec}(\B) \in \Re^{\tau L}$ is the ultra-high-dimensional
coefficient vector.
Instead of fitting this model directly, however, we propose to approximate this matrix GLM with a series of local surrogates, 
$g(\boldsymbol{\mu}_l)$, that we blend together in an ensemble through spatial regularization.  The key assumption we make is that conditional on the spatial smoothness of our parameters, the parameters of each local model are such that local models are approximately independent.  Beyond this, however, we will see that our local models offer several advantages in terms of mathematical complexity, computational complexity and statistical efficiency.  
 
 To understand some of the difference between the matrix GLM and our Local-Aggregate model as well as the advantages of our approach, let us compare methods for the simple case of regression with squared error loss.  Here, we will assume we are working with spatio-temporal neuroimaging data and hence employ the same temporal and spatial smoothing penalties.  Consider the matrix GLM:
\begin{equation}
\label{eq1}
\min_{\B} \| \y - \X_{(1)}vec(\B)\|^2 + \sum_{l=1}^L \lambda_{loc}\bbeta_l^T \Omeg\bbeta_l + \lambda_{agg}tr(\B\G\B^T).
\end{equation}
To solve this multivariate model, we rewrite the problem as
\[
\min_{\B} \| \y - \X_{(1)}\Pb \B \q\|^2 + \lambda_{loc} tr(\B^T\Omeg \B) + \lambda_{agg}tr(\B\G\B^T),
\]
where  $\q = \begin{pmatrix} 1\\ 0 \\ 0 \end{pmatrix}$
, $\Pb = \begin{pmatrix} \I \\ \I_{1,2} \\ \vdots \\ \I_{1,L}\end{pmatrix} \in \Re^{\tau L \times \tau}$, $\I_{1,l}, l = 2,\ldots,L$ are identity matrix with the 1st and $l^{th}$ column switched. Setting the gradient of the multivariate loss function to $0$, we have
\[
\Pb^T \X_{(1)}^T(\y - \X_{(1)}\Pb\B)\q^T+ \lambda_{loc} \Omeg \B + \lambda_{agg} \B\G = \0
\]
of the form "$ \A\X\B + \C\X + \X\D = \cc$", which does not have an analytical solution and requires a semidefinite quadratic linear programming solver.  On the other hand, our Local-Aggregate model for this problem takes the following form:
\begin{equation}
\label{eq2}
\min_{\B} \sum_{l=1}^L\| \y - \X_l \bbeta_l\|^2 + \sum_{l=1}^L \lambda_{loc}\bbeta_l^T \Omeg\bbeta_l + \lambda_{agg}tr(\B\G\B^T)
\end{equation}
Thus we break down the multivariate model into local models that are mathematically much easier to solve and have analytical solutions: $\bbeta_l = (\X_l^T\X_l + \lambda_{loc}\Omeg)^{-1} \X_l \y$.
Hence, our Local-Aggregate method is mathematically much easier to solve than the multivariate model.

As discussed in Section 2, we can distribute computations, data storage and memory for our 
 Local-Aggregate model is distributable because our loss is location-seperable: $\ell(\Xcal,\B) = \sum_{l=1}^L \| \y - \X_l \bbeta_l\|_2^2
= \sum_{i=1}^n \sum_{l=1}^L (y_i - \sum_{t=1}^\tau x_{itl}\beta_{tl})^2$. Thus computationally, our model scales well for big data that is collected and stored in a distributed manner.  
On the other hand, optimization of the multivariate model cannot be parallelized due to the order of summations in the loss function: $\ell (\Xcal,\B) = \| \y - \X_{(1)}\mathrm{vec}(\B) \|_2^2 = \sum_{i = 1}^n (y_i - \sum_{t = 1}^\tau \sum_{l = 1}^L x_{itl} \beta_{tl} )^2.$   Solving the matrix GLM is then computationally intractable for big data that are too large to load into the memory of one single computer.

The multivariate model vectorizes the $\B$ coefficient matrix which leads to an ultra-high-dimensional problem where $ n \ll \tau L$. Because of this, it is not only computationally burdensome, but it is also statistically inefficient to solve.  As our Local-Aggregate method breaks down the problem into much more manageable local models of size $n \times \tau$. The sample size complexity $n$ relative to $\tau$ versus $n$ relative to $\tau L$, is greatly reduces.  We then expect higher statistical efficiency in estimating the local models leading to overall improved predictive accuracy with respect to the multivariate model.

We investigate this assertion in a small simulated example. Here, our setup is the same as the that in the regression simulation example presented in Section 5 except we replace the sparse spatial signals with smooth sinusoidal spatial signals, and we use a smaller problem size with $n = 50, \tau = 50$.  We compare the different scenarios of $L = 25, 64, 100$, and use the MSE as the criterion for the prediction error.  We use the cvx solver for the optimization of the matrix GLM in \eqref{eq1}.

As indicated in Table~\ref{table:mult}, our Local-Aggregate model has much better prediction error than the matrix GLM in all scenarios and offers substantial improvements when $L$ is larger.  We expect that when $L$ is in the tens of thousands as in neuroimaging, our method will yield far better performances as the sample complexity of our local models is constant at $\tau$ compared to $L\tau$ for the matrix GLM.

\begin{table}
\label{table:mult}
\centering
{\small
\begin{tabular}{ l l l}
\hline
 \rowcolor{Tgray} &Method& Prediction Error  \\
 \hline
L = 25&Loc-Agg & 0.3864 (0.0066)  \\ 
&Matrix GLM& 0.6494 (0.0178)  \\ 
&Local& 0.7017 (0.0075)\\ 
\hline 
L = 100&Loc-Agg &1.0408(0.0085)  \\ 
&Matrix GLM& 2.1533 (0.0245)  \\ 
&Local& 1.3002 (0.0098)\\ 

\hline
L = 144 &Loc-Agg &1.2348 (0.0136)  \\ 
&Matrix GLM& 2.2141 (0.0231)  \\ 
&Local &1.4677 (0.0147) \\ 

\hline
\end{tabular}
\caption{Comparisons of prediction accuracy for the Local-Aggregate model (Eq C.2), the matrix GLM (Eq C.1), and the Local model (Eq C.3) on the spatial-temporal regression simulation. Here $n = 50, \tau = 50$ and we vary the number of locations $L$.}
}
\end{table}
One may wonder that the good performance of the Local-Aggregate method mainly comes from using the separate local loss functions $\sum_{l=1}^L \| \y -\X_l \bbeta_l\|_2^2$ to approximate  $\| \y - \X_{(1)} \text{vec}(\B)\|_2^2$, but less from the aggregating penalty, $P_{agg}(\B,\G)$, as the Matrix GLM also uses the aggregating penalty.  
We propose to investigate this question with comparison of our method with the Local model with just the local loss functions and local penalties:
\begin{equation}
\label{eq3}
\min_{\B} \sum_l \| \y - \X_l \bbeta_l\|_2^2+  \sum_{l=1}^L \lambda_{loc}P_l (\bbeta_l).
\end{equation}
We can see that although the separate loss functions with local penalties alone have better performance compared with the multivariate model when $L$ is large, its performance is not as good as our Local-Aggregate method where both the proposed loss function and aggregating penalties are applied. 

In conclusion, compared with the multivariate model, our Local-Aggregate modeling framework is (1) mathematically much easier to solve, (2) computationally more efficient as is conducive to distributed optimization and (3) yields improved statistical efficiency.

\section{Convergence Accelerators for Local-Aggregate ADMM Algorithm}

The stopping criterion of ADMM algorithms typically depends on the norm of the primal and
dual residuals \citep{boyd2011distributed,wahlberg2012admm,annergren2012admm}.  We take the stopping criterion to be $\|\rr\| < \epsilon^{pr}$ and $\|\s \|< \epsilon^{dual}$ for some $\epsilon^{pr}, \epsilon^{dual}>0$ as suggested in \citet{boyd2011distributed}.  Additionally, the parameter $\rho > 0$ is taken as
fixed throughout the algorithm as recommended by
\citet{boyd2011distributed}. Under the assumptions of convex GLM loss and penalties and fixed $\rho$, the convergence of the algorithm is ensured, and the convergence rate is $\mathcal{O}(1/k)$ \citep{luo2012linear,deng2012global,goldstein2012fast}, where k is the iteration number.  This convergence speed is still too slow for the purpose of big data modeling, and can be very computationally expensive, for example, it would require at least 1000 iterations even for a rough accuracy of 1e-3.  Therefore we need to further speed up the convergence of the algorithm, and cut down the number of iterations to within hundreds. 

The convergence of the ADMM can be improved by adapting 
the parameter $\rho$ to the magnitude of primal and dual residuals
across iterations and thereby reducing the dependency on the initial choice of the penalty parameter.  
\citet{boyd2011distributed,he2000alternating} have proposed the following adaptive penalty parameter scheme that tries to keep the norms of primal  redisudal ($\rr = \B-\Z$) and dual residual ($\s^k = \Z^k- \Z^{k-1}$) within a certain factor of each other as they both converge to zero:
\[
 \rho^{k+1} = \begin{dcases*}
        \tau^{incr}\rho^k  & if $\|\rr^k\|_2 > \mu \|\s^k\|_2$ \\
        \rho^k / \tau^{decr}& if $\|\s^k\|_2 > \mu \|\rr^k\|_2$ \\
        \rho^k  & otherwise.      
        \end{dcases*},
\]
where the two multipliers $\tau^{incr}, \tau^{decr}$ control the acceleration of convergence: the larger $\tau^{incr}, \tau^{decr}$ are, the faster the ADMM converges.  One common choice may be $\tau^{incr} = \tau^{decr} = 2$ and $ \mu = 10$.  Other approaches include monotonically increasing or decreasing $\rho$ updates \citep{he2000alternating}, which depend strongly on the starting values of $\rho$; the self-adaptive updating scheme automatically adjust for the starting value of $\rho$.   However, these schemes do not respect the spatio-temporal structure of the data for our particular problem. Additionally, slow convergence at one time point or location may encumber the overall convergence speed of the entire procedure. 

We propose a new technique with self-adaptive penalty parameters that fully respect the complex data structure.
We can further speed up the convergence by taking into account of the temporal structure of the data: instead of using one scalar-value, $\rho$, for all time points and locations, we use a different $\rho_t$ at each time point.  The temporal measurements of neuroimaging data exhibit strong spikes in response to stimuli.  Hence given the same step size $\rho$, the convergence rate for these spiked activation periods is slower than at other times because of larger residuals.  By forcing the convergence speed to be relatively faster at time points with larger primal residuals, and relatively slower at time points with larger dual residuals, we obtain more balanced convergence speeds over all time points:
\[
 \rho_t^{k+1} = \begin{dcases*}
        \tau^{incr}\rho_t^k  & if $\|\rr_t^k\|_2 > \mu \|\s_t^k\|_2$ \\
        \rho_t^k / \tau^{decr}& if $\|\s_t^k\|_2 > \mu \|\rr_t^k\|_2$ \\
        \rho_t^k  & otherwise,      
        \end{dcases*}
\]
where $\|\rr_t\|_2^2 = \sum_{l=1}^{L}\rr_{tl}^2$ is the primal residual at time point $t$,  $\|\s_t\|_2^2 = \sum_{l=1}^{L}\s_{tl}^2$ is the dual residual at time point $t$.  We switch to a fixed $\rho$ updating scheme, i.e., $ \tau^{incr} =  \tau^{decr}  = 1$ after $K = 1000$ iterations.

In order to coordinate convergence speeds at different time points, we propose a new Augmented Lagrangian for the Local-Aggregate modeling framework:
\begin{equation}
\label{lagrangian_accl}
\mathcal{L}_{\rrho}(\B, \Z, \U) = \sum_{l = 1}^{L} \big[ r(\y;
       \X_{l} \bbeta_{l} ) + \lambda_{loc} P_{loc}( \bbeta_{l}) \big] 
      +\lambda_{agg}P_{agg}(\Z,\G) + \sum_{l = 1}^{L} \frac{1}{2}\| \bbeta_l-\z_l +\uu_l \|_{\rrho}^2,
\end{equation}
where $\| \bbeta_l-\z_l +\uu_l \|_{\rrho}^2 =  (\bbeta_l-\z_l +\uu_l)^T diag(\rrho) ( \bbeta_l-\z_l +\uu_l)$.

The $\Z$-subproblem, therefore, no longer has a simple least squares solution in the vector $\rrho$ case.
\begin{lemma}
The solution of the $\Z$-updates of the Local-Aggregate ADMM is equivalent to the solution of \[diag(\rrho)\Z + \Z(2\gamma\G) = diag(\rrho)(\B+\U),\]which is in the form of the Sylvester equation $\A\X+\X\B = \C$.
\end{lemma}

 The Sylvester equation does not have an analytical solution when the rank of either $\A$ or $\B$ is greater than 1. Since there is no analytical solution \citep{sorensen2003direct}, the MATLAB numerical solver {\em lyap} is used.

\begin{figure}[htbp] 
   \centering
   \includegraphics[width=4in]{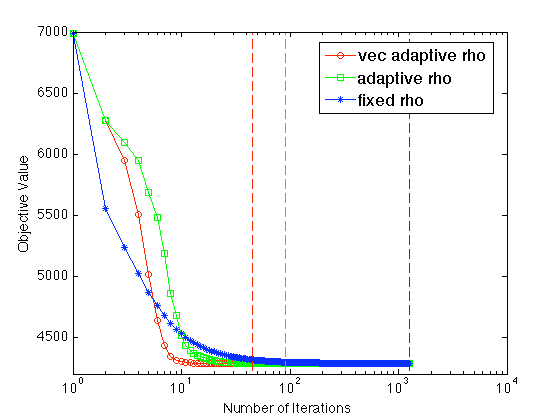} 
   \caption{Comparisons of different updating schemes for $\rrho$. The number of iterations till convergence is 45, 91, and 1270 for self-adaptive vectorized $\rrho$, self-adaptive scalar $\rho$ and fixed scalar $\rho$, respectively.}
   \label{convergence}
\end{figure}

The Local-Aggregate ADMM with the new self-adaptive vectorized $\rrho$ penalty parameter is guaranteed to converge, as shown in {\bf Theorem   C.2.} We also provide a proof of this theorem in Section E.

As shown in Figure~\ref{convergence}, our self-adaptive vectorized $\rrho$ updating scheme yields the fastest monotonic convergence (approximately half of the number of iterations of scalar self-adaptive $\rho$ scheme). The self-adaptive scalar $\rho$ updating scheme is faster in the initial fast-decay stage (the first 5 iterations), and then slows down due to the difference in convergence speeds among time points.  The fixed $\rho$ scheme is the slowest amongst all, and depends on the choice of $\rho$ value.  Hence, our self-adaptive vectorized $\rrho$ updating scheme yields the best convergence performance due to its consideration of the spatio-temporal structure of the data.

To prove convergence of our vector adaptive $\rrho$, we write it in general form:
\[
\begin{aligned}
   &\underset{\B, \Z}{\text{minimize}} 
   && f(\B) + g(\Z) \\
   & \text{subject to}
   && \B =\Z,
\end{aligned}
\]
where 
 \[
f(\B) =  \sum_{l = 1}^{L}\left[ r(\y;
      \alpha_{l} + \X_{l} \bbeta_{l} ) + \lambda_{loc}P_{loc}(\bbeta_l)  \right], \quad \quad
      g(\Z) = \lambda_{agg} \text{ tr}(\Z \G \Z^T).
   \]
   
   \addtocounter{algorithm}{+2}

Algorithm~\ref{alg:adap}  then can be put into the following form, letting $ \HH = diag(\rrho) $, $f(\Z)$ and $g(\Z)$ are proper, closed, convex functions. 

\begin{algorithm}
\caption{Local-Aggregate Algorithm with Self-Adaptive Vector $\rrho$}
\label{alg:adap}
\begin{align}
&\label{f1}1.  \B^{k+1} \in  \argmin_{\B} \quad f(\B) + \sum_{l = 1}^L[ (\vv_l^{k})^T\HH^k\bbeta_l + \frac{1}{2}\|\bbeta_l - \z_l^k\|_{\HH^k}^2 ]\\
 &\label{f2} 2. \Z^{k+1}=  \argmin_{\Z} \quad g(\Z) +\sum_{l = 1}^L[ -(\vv_l^{k})^T\HH^k\z_l +\frac{1}{2}\|\bbeta_l^{k+1} - \z_l\|_{\HH^k}^2 ]\\
 &\label{f3}3. \V^{k+1} = \V^k + (\B^{k+1}- \Z^{k+1})  \\
 & \nonumber \text{4. Update} \HH^k
\end{align}
\end{algorithm}

We first make a basic assumption so that the problem has an optimal solution with finite objective value.
\newline
\textbf{Assumption 1.} Problem~\ref{f1} admits a Lagrangian saddle point, i.e., there exist $\B^* \in \text{dom } f, \Z^* \in \text{dom } g$ and $\V^*$ such that $\X^* = \B^*$ and $f(\B^*) + g(\Z^*) \leq f(\B) + g(\Z) + \sum_{l = 1}^L \vv_l^{*T}(\bbeta_l -\z_l), \forall \B \in \text{dom } f, \forall \Z \in \text{dom }g.$
\newline
\textbf{Assumption 2.} Problems~\ref{f1} and \ref{f2} are solvable.
\newline
\textbf{Assumption 3.} The eigenvalues of the semi-positive-definite matrix $\HH^k$ are uniformly bounded from below away from zero, and with finitely many exceptions, the eigenvalues of $\HH^k - \HH^{k+1}$ are nonnegative.

The existence of a saddle point is proved in \citet{gabay1976dual} and the two subproblems are solvable since $f$ and $g$ in our Local-Aggregate framework are proper convex functions.  Our updating scheme for $\HH$ is such that after $K = 1000$ iterations, $\HH^k - \HH^{k+1} = 0$, and we switch to an constant $\HH$ updating scheme so that Assumption 3 is always satisfied.  Following from Theorem 2.1. in \citet{kontogiorgis1998variable}, we have the following theorem:   
\begin{theorem} 
\label{converg_vec_rho}If assumptions 1-3 hold, then for any sequence of iterates $ \{\B^k, \Z^k, \V^k, \HH^k \}$ produced by the alternating direction method :
\begin{itemize}
\item[(i)] $\{\B^k, \Z^k\}$ converges, and the limit satisfies the constraints of \eqref{f1}
\item[(ii)]$\{f(\X^k), g(\Z^k)\}$ converges to the optimal value of the objective function for problem \eqref{f1}. 
\item[(iii)]$\{\HH^k\V^k\}$ converges to an optimal dual multiplier for problem \eqref{f1}.
\item[(iv)]Any minimizers of problems of the form \eqref{f2} and \eqref{f3} in which $\HH^k\V^k$, $\B^{k+1}$ and $\Z^k$ are fixed at their limit values are optimal for problem \eqref{f1}.
\end{itemize}
\end{theorem}

Since Assumptions 1, 2 and 3 are all satisfied, our Local-Aggregate algorithm converges.  Although there is currently no theoretical proof of a faster convergence speed for the adaptive penalty parameter, the convergence speed is empirically improved as illustrated in Figure~\ref{convergence}.

\section{Classification Simulation}
We will further explore the performance of Local-Aggregate modeling framework for classification problems with local Logistic loss functions.  
We use the same simulation setup as previously described in the regression simulations in Section 5, except we generate the binary response variable $\y_i 
\sim Bernoulli( \frac{1}{1+e^{-\X_{(1)}vec(\B)}})$.  We compare our method to the linear support vector machine (SVM), the nuclear-norm regularized tensor logistic regression method \citep{zhou2013tensor}, the logistic Lasso, and logistic Elastic Net. We will also explore the effect of the number of locations and the rank of the true signal. 
\begin{table}
\label{table2}
{\scriptsize
\begin{center}
\caption{\small Classification results for varying the (top) Number of Locations, with $n = 200$, $\tau = 200$, and $rank(\B_o) = 2$, and (bottom) the rank of true signal $\B_o$, with $L = 100$.}
\begin{tabular}{ l  l  l*{4}l }
\hline
&   &  Method & Prediction Error  & MSE(B) & FPR & TPR\\
\hline
\hline
L& 25 &Loc-Agg &24.26\%(0.81\%)&0.7973(0.0140)&72.14\%(4.25\%)&98.50\%(0.85\%)\\
 & & LASSO   &33.18\%(1.01\%)&1.1807(0.0098)&55.14\%(3.37\%) &89.00\%(4.30\%)\\
 & &Elastic Net    &28.22\%(0.89\%)&0.9723(0.0090)&89.14\%(2.77\%)&96.00\%(2.80\%)\\
&&SVM &25.94\%(0.68\%)&1.6177(0.0105)& / & /  \\
 \rowcolor{Tgray}&&Tensor Reg.   &23.66\%(0.68\%)&0.9264(0.0186)& / & / \\
\hline
 \rowcolor{Tgray}& 100 &Loc-Agg &23.78\%(0.74\%)&0.7806(0.0161)&28.72\%(3.90\%)&97.25\%(0.06\%)\\
  && LASSO   &40.34\%(0.80\%)&1.1085(0.0095)&15.24\%(1.50\%) &60.75\%(4.78\%)\\
  &&Elastic Net    &36.70\%(0.86\%)&1.0599(0.0093)&45.28\%(3.40\%)&84.50\%(4.69\%)\\
&&SVM &34.02\%(0.74\%)&1.6780(0.0065)& / & /  \\
&&Tensor Reg.   &30.54\%(0.83\%)&0.9861(0.0125)& / & / \\
\hline
  \rowcolor{Tgray}&144 &Loc-Agg &24.98\%(0.90\%)&0.8010(0.0148)&21.31\%(3.11\%)&91.00\%(2.42\%)\\
  && LASSO   &42.82\%(0.93\%)&1.0798(0.0114)&8.01\%(1.22\%) &37.25\%(4.90\%)\\
  &&Elastic Net    &38.22\%(1.05\%)&1.0523(0.0076)&40.40\%(2.83\%)&84.50\%(4.55\%)\\
&&SVM &35.58\%(0.73\%)&1.6611(0.0068)& / & /  \\
&&Tensor Reg.   &34.36\%(0.94\%)&1.0365(0.0093)& / & / \\
\hline
\hline
 \rowcolor{Tgray} Rank ($\B_o$)&2&Loc-Agg &23.78\%(0.74\%)&0.7806(0.0161)&28.72\%(3.90\%)&97.25\%(0.06\%)\\
  && LASSO   &40.34\%(0.80\%)&1.1085(0.0095)&15.24\%(1.50\%) &60.75\%(4.78\%)\\
  &&Elastic Net    &36.70\%(0.86\%)&1.0599(0.0093)&45.28\%(3.40\%)&84.50\%(4.69\%)\\
&&SVM &34.02\%(0.74\%)&1.6780(0.0065)& / & /  \\
&&Tensor Reg.   &30.54\%(0.83\%)&0.9861(0.0125)& / & / \\
\hline
 \rowcolor{Tgray}& 8 &Loc-Agg &22.58\%(0.79\%)&0.7743(0.0178)&19.43\%(3.59\%)&91.25\%(2.15\%)\\
 & & LASSO   &39.00\%(0.97\%)&1.1043(0.0091)&15.57\%(1.53\%) &60.25\%(4.88\%)\\
 & &Elastic Net    &34.60\%(0.95\%)&1.0221(0.0099)&48.00\%(2.99\%)&92.75\%(3.07\%)\\
&&SVM &32.44\%(0.58\%)&1.6916(0.0067)& / & /  \\
&&Tensor Reg.   &28.74\%(0.74\%)&0.9594(0.0116)& / & / \\
\hline
\end{tabular}
\end{center}
}
\end{table}

\vspace*{.1in}
\noindent{\em E.1  Number of Locations.  }
We employ the same simulation setup with different numbers of locations: $L = 25, 100, 144$. Table~\ref{table2} (top) suggests that as the number of locations increases, the prediction accuracy of Local-Aggregate method remains approximately the same while other methods' performance deteriorates. 
The Local-Aggregate method achieves the best signal recovery as well. The Lasso method over-sparsifies the coefficients due the high dimensionally of the data. We can see that the results are less sensitive to the influence of the number of locations than the regression simulation results, since the response variable only depends on the sign of the linear predictor $\X_{(1)}vec(\B)$. 
\vspace*{.1in}

\noindent{\em E.2  Signal Complexity.  }  
Now we investigate the behavior of our method under different ranks of the signals given the same number of locations and amount of spatio-temporal correlation. Table~\ref{table2} (bottom) suggests that the larger rank of the spatial-temporal signal, the better the detection of the non-zero locations; other regularized regression methods are insensitive to the change of rank of $\B_o$.

\section{Proofs}

%
\noindent {\bf Theorem 3.1.}
\begin{proof}
\citet{boyd2011distributed} and \citet{mota2011proof} together gives a proof of the ADMM algorithm.
Let 
\[
\begin{split}
&f(\B) = \sum_{l = 1}^L r(\y;\X_l \bbeta_l ) + \lambda_{loc}P_{loc}(\bbeta_l), \\
&g(\Z) =\lambda_{agg}P_{agg}(\Z,\G),\\
\end{split}
\]
we can rewrite the Local-Aggregate Framework as
\begin{equation}
\label{admm2}
   \underset{\B,\Z}{\text{minimize}}  \quad f(\B) + g(\Z) \quad \text{subject to} \quad \B = \Z.
\end{equation}
The augmented Lagrangian of \eqref{admm2} is
\[
\mathcal{L}_{\rho}(\B, \Z, \V) = f(\B) + g(\Z) ++ \sum_{l = 1}^{L} [\vv_l^T (\bbeta_l- \z_l) + \frac{\rho}{2}\| \bbeta_l-\z_l \|_2^2].
\]
\citet{boyd2011distributed}  shows that if $f$ and $g$ are closed, proper, and convex, and the unaugmented Lagrangian $\mathcal{L}_0$ has a saddle point $(\B^*, \Z^*, \V^*)$, then we have primal residual convergence, i.e., $\rr^k = \B^k- \Z^k \rightarrow 0$ as $ k \rightarrow \infty$, dual residual convergence, i.e., $\s^k = \rho(\Z^k-\Z^{k-1})\rightarrow 0$, and also objective convergence, i.e., $p^k \rightarrow p^*$, where $p^k  = f(\B^k) + g(\Z^k)$.   \citet{mota2011proof} further shows that the dual variable $\V^k$ converges to the dual optimal point $\V^*$.
The only difference from the proof in \citet{boyd2011distributed} and \citet{mota2011proof} is to
 define a different Lyapunov function of the algorithm:
\[
Q^k =\sum_{l = 1}^L[ (1/\rho) \| \vv_l^k - \vv_l^* \|_2^2 +\rho \| \z_l^k - \z_l^*\|_2^2 ] .
\]

The convergence results of relies on proving the following three key inequalities, and the proof details are the same as in \citet{boyd2011distributed} and \citet{mota2011proof} except for the different $Q$ function.
\begin{equation}
\label{ineq1}
Q^{k+1} \leq Q^k - \sum_{l= 1}^L [ \rho \| \bbeta_l^{k+1} - \z_l^{k+1}\|_2^2 + \rho \| \z_l^{k+1} - \z_l^k\|_2^2].
\end{equation}
\begin{equation}
\label{ineq2}	
p^{k+1}-p^* \leq \sum_{l = 1}^L[ -(\vv^{k+1})^T(\bbeta_l^{k+1} - \z_l^{k+1}) - \rho(\z_l^{k+1}-\z_l^k)^T(\bbeta_l^{k+1} - \z_l^{k+1} + \z^{k+1}-\z^*)],
\end{equation}
\begin{equation}
\label{ineq3}
p^* - p^{k+1} \leq \sum_{l = 1}^L (\vv_l^*)^T(\bbeta_l^{k+1} - \z_l^{k+1}).
\end{equation}

\end{proof}

\noindent {\bf Lemma   B.1.}
\begin{proof}
We need to reparametrize the problem \eqref{loc_agg_eeg_orig} by 
letting $\widetilde{\x}_{il} = [1 \x_{il}] \in \Re^{(\tau+1)}$, $\widetilde{\bbeta_l} = [\alpha_l; \bbeta_l] \in \Re^{\tau+1}$, $\Sb = 
\begin{pmatrix}
  0 &  \cdots &0 \\
  \bf{0} &  \I_{\tau}
 \end{pmatrix} \in \Re^{(\tau+1)\times(\tau+1)}$ , $\widetilde{\Omeg} = 
 \begin{pmatrix}
  0 &  \cdots &0 \\
  \bf{0} &  \Omeg 
 \end{pmatrix}\in \Re^{(\tau+1)\times(\tau+1)}$, problem \eqref{loc_agg_eeg_orig} becomes
 
 \begin{equation}
\label{loc_agg_eeg_orig_reparam}
\begin{aligned}
& \underset{ \widetilde{\B}}{\text{minimize}}  \sum_{l = 1}^{L}[  \sum_{i = 1}^n\{-\y_i(\widetilde{\x}_{il}^T \widetilde{\bbeta}_l) + \log(1 + e^{\widetilde{\x}_{il}^T\widetilde{\bbeta}_l})  \}+ 
\lambda_{loc}^{sm}   \widetilde{\bbeta}_{l}^{T}  \widetilde{\Omeg} \widetilde{\bbeta}_{l}\\
     & \quad \quad \quad +      \lambda_{loc}^{sp} \|
      \Sb \widetilde{\bbeta}_{l} \|_{2} ] 
        + \lambda_{agg} \text{ tr}((\Sb \widetilde{\B})\G (\Sb \widetilde{\B})^T). \\
\end{aligned}
\end{equation}
Let $\widetilde{\Z }= \Sb \widetilde{\B}$, we can set up the Local-Aggregate ADMM as
 
 \begin{equation}
\label{loc_agg_eeg_reparam}
\begin{aligned}
&\underset{\widetilde{\B},\widetilde{ \Z}}{\text{minimize}} \quad
 \sum_{l = 1}^{L}[\sum_{i = 1}^n\{-\y_i(\widetilde{\x}_{il}^T \widetilde{\bbeta}_l) + \log(1 + e^{\widetilde{\x}_{il}^T\widetilde{\bbeta}_l})  \}+ \lambda_{loc}^{sm}\widetilde{\bbeta}_{l}^{T}\widetilde{\Omeg} \widetilde{\bbeta}_{l}\\
&  \quad \quad \quad+\lambda_{loc}^{sp} \|\Sb \widetilde{\bbeta_l} \|_2 ]+ \lambda_{agg} \text{tr}(\widetilde{\Z} \G \widetilde{\Z}^T)\\
&  \text{subject to}\quad  \widetilde{\Z} =\Sb \widetilde{\B}.
\end{aligned}
\end{equation}

After rename the variables without the $ \text{ }\widetilde{}  \text{ }$ sign, we have problem \eqref{loc_agg_eeg} equivalent to \eqref{loc_agg_eeg_orig}.

\end{proof}

\noindent {\bf Lemma   B.2. }
\begin{proof}
 \[
 \begin{split}
& \frac{\partial \text{tr}(\Z\G\Z^T) + \sum_{l = 1}^{L} \frac{\rho}{2}\| \z_l - \Sb\bbeta_l^{k+1} - \uu_l^{k} \|_2^2}{\partial \Z} = 2 \lambda_{agg} \Z \G + \rho(\Z-\Sb\B^{k+1}-\U^{k}) = 0\\
& \Z(2 \lambda_{agg} \G + \rho \I_L) = \rho(\Sb\B^{k+1}+\U^k).
 \end{split}
 \]
 \end{proof}

\noindent {\bf Lemma   C.1. }
\begin{proof}
\[
\frac{\partial(\gamma \text{ tr}(\Z\G\Z^T)+  \frac{1}{2}\sum_{l=1}^L \|\z_l - \bbeta_l^{k+1}-\uu_l^k\|_{\rrho}^2)}{\partial \Z} = 2\gamma\Z\G + diag(\rrho)(\Z-\B-\U),
\]
Setting the gradient to 0 and rearranging terms, we obtain
The $\Z$-subproblem is thus reduced to solving the Sylvester equation, where $\A = diag(\rrho)$, $\B = 2\gamma\G$, and $\C = diag(\rrho)(\B+\U)$.
\end{proof}

\vspace*{.1in}

\noindent {\bf Theorom   C.2. }
\begin{proof}
We will sketch the outline the proof as a collection of lemmas adapted from \citet{kontogiorgis1998variable}, go over the major results and skip the details of the proofs.

We begin the proof by showing in {\bf Lemma   E.1} that the iterates $\{(\B^k, \Z^k, \V^k)\}$ are bounded. Then we establish that $\{f(\B^k) + g(\Z^k)\}$ converges to the optimal objective value of the problem \eqref{admm2} in {\bf Lemma   E.2}. We then show {\bf Lemma   E.3} that $\{\B^k\}$, $\{ \Z^k\}$, and $\{\HH^k\uu^k\}$ converge, and that the limit of $\{\HH^k\uu^k\}$ is an optimal dual for \eqref{admm2}. Finally, {\bf Lemma   E.4} shows how to obtain a primal solution for \eqref{admm2} by solving two minimization subproblems, using the limits of $\{\B^k\}$, $\{ \Z^k\}$, and $\{\HH^k\uu^k\}$ as fixed terms in the objective.

\begin{lemma}
Let the assumptions of Theorem   C.2 hold, then $\{\B^{k}\}$, $\{\Z^{k}\}$ and $\{\V^{k}\}$ are bounded.
\end{lemma}

\begin{lemma}
Let the assumptions of Theorem   C.2 hold, then
\newline (i) $\{\V^{k+1}-\V^k \rightarrow 0\}$, $\{\Z^{k+1}-\Z^k\}\rightarrow 0$, $\{\B^k - \Z^k\}\rightarrow 0$.
\newline (ii) $\{f(\B^k) + g(\Z^k)\}$ converges to the optimal objective value for problem \eqref{admm2}
\end{lemma}

\begin{lemma}
Let the assumptions of Theorem   C.2 hold, then
\newline (i)  $\{\B^{k}\}$, $\{\Z^{k}\}$ and $\{\V^{k}\}$ converge.
\newline (ii) The limit of $\{\HH^k\V^k\}$ is an optimal dual for problem \eqref{admm2}.
\end{lemma}

\begin{lemma}
Let $\{\Z^k, \HH^k\V^k\}$ converge to $(\bar{\z}, \bar{\vv})$. Let $\widetilde{\B}$ solve
\[
\underset{\B}{\text{minimize}}\quad f(\B) + \sum_{l=1}^L\bar{\vv_l}^T\bbeta_l + \phi_1(\B-\bar{\z}),
\]
and let $\widetilde{\Z}$ solve
\[
\underset{\Z}{\text{minimize}} \quad g(\Z) - \sum_{l=1}^L\bar{\vv_l}^T\z_l + \phi_2(\bar{\z}-\Z),
\]
in which $\phi_1$ and $\phi_2$ are continuous positive definite functions. Then $(\bar{\B}, \bar{\Z})$ solves \eqref{admm2}.
\end{lemma}

Part (i) of the theorem can be proven by combining part (i) of {\bf Lemma   E.2} and part (i) of {\bf Lemma   E.3}. Part (ii) of the theorem is proven in part (ii) of {\bf Lemma   E.2} and part (iii) is proven in part (ii) of {\bf Lemma   E.3}. Part (iv) is a special case of {\bf Lemma   E.4}.

The proof of {\bf Theorem   C.2.} follows from \citet{kontogiorgis1998variable} with some minor modification:
\newline
 - the Augmented Lagrangian of the problem \eqref{admm2} :
\[
\mathcal{L}(\B,\Z,\U) = f(\B)+g(\Z) + \sum_{l = 1}^L[ (\vv_l)^T\HH(\bbeta_l-\z_l) +\frac{1}{2}\|\bbeta_l - \z_l\|_{\HH}^2 ],
\]
\newline
- when proving {\bf Lemma   E.1.} (Lemma 2.5 in \citet{kontogiorgis1998variable}), we use different trick functions
\[
J_1(\B) = f(\B), \quad J_2(\B) = \sum_{l = 1}^L[ (\vv_l^{k})^T\HH^k\bbeta_l + \frac{1}{2}\|\bbeta_l- \z_l^k\|_{\HH^k}^2 ],
\]
\[
J_1(\Z) = g(\Z), \quad J_2(\Z) = \sum_{l = 1}^L[ -(\vv_l^{k})^T\HH^k\z_l +\frac{1}{2}\|\bbeta_l^{k+1} - \z_l\|_{\HH^k}^2 ].
\]

\end{proof}

{\singlespacing
 \bibliographystyle{spmpsci}
 {\footnotesize\bibliography{loc-agg}}}

\end{document}